\documentclass[11pt,prd,superscriptaddress,nofootinbib]{revtex4}

\usepackage[usenames,dvipsnames]{xcolor}

\def\be{\begin{equation}}
\def\ee{\end{equation}}
\def\ba{\begin{array}}
\def\ea{\end{array}}
\def\bea{\begin{eqnarray}}
\def\eea{\end{eqnarray}}
\def\nn{\nonumber\\}

\def\ph{\phi}

\def\m{\mu}

\def\r{\rho}

\def\o{\omega}

\def\pa{\partial}

\def\ls{\left(}

\def\rs{\right)}

\begin{document}

\title{Meson effective mass in isospin medium in hard-wall AdS/QCD model}
\author{Shahin Mamedov}
\email{sh.mamedov62@gmail.com}
\affiliation{Department of Physics, Gazi University, Teknikokullar, 06100 Ankara, Turkey}
\affiliation{Institute for Physical Problems, Baku State University, Z.Khalilov 23, AZ-1148, Baku, Azerbaijan}
\affiliation{Institute of Physics, Azerbaijan National Academy of Sciences, H.Javid ave. 33, AZ-1143, Baku, Azerbaijan}

\maketitle

\centerline{\bf Abstract} \vskip 4mm
We study a mass splitting of light vector, axial-vector and pseudoscalar mesons in isospin medium in the framework of hard-wall model. We write an effective mass definition for the interacting gauge fields and scalar field introduced in gauge field theory in the bulk of AdS space-time.  Relying on holographic duality we obtain a formula for the effective mass of a boundary meson in terms of derivative operator over the extra bulk coordinate. The effective mass found in this way coincides with the one obtained from finding of poles of the two-point correlation function. In order to avoid introducing distinguished infrared boundaries in the quantisation formula for the different mesons from the same isotriplet we introduce extra action terms at this boundary, which reduces distinguished values of this boundary to the same value. Profile function solutions and effective mass expressions were found for the in-medium $\rho$, $a_1$ and $\pi$ mesons.
\vspace{1cm}

\section{Introduction}
Isospin medium is the simplification of the dense nuclear medium, where the net baryon
charge of the medium is taken zero while its isospin chemical potential remains non-zero. Such a simplified model in QCD was introduced in \cite{1}. The AdS/CFT correspondence conjecture developed in \cite{2,3,4,5,6} further was applied to QCD problems and AdS/QCD models were constructed for the mesons \cite{7,8,9,10,11,12,13,14}. Nucleons were included into AdS/QCD models in \cite{15,16}. The AdS/QCD idea was extended to the finite temperature case in  \cite{17,18,19,20,21,22,23,24,25,26} and a great number of works were devoted to a holographic description of the quark-gluon plasma and dense nuclear matter \cite{27,28,29}. In the framework of holographic QCD the studies in the dense nucleon and isospin mediums turned out as effective in a top-down approach \cite{30,31,32,33} as in a bottom-up one \cite{34,35,36,37,38,39}.

In holographic QCD the two phases of nuclear matter - the confining and deconfining ones, in the dual gravity theory side are described by the different metrics. In the bottom-up approach, when the quark matter is absent,  the deconfining phase at the gravity side is described  by the Schwarzschild AdS black hole (SAdS BH) metric \cite{17}. The confined phase of the nuclear matter at the low temperature limit in the dual gravity theory is described by the thermal AdS space (tAdS) \cite{40,41}. In \cite{42} it was shown on the background geometry of the dual gravity for the confinement phase containing the quark matter fields as well. This metric is named the thermal charged AdS space (tcAdS), and it can be obtained from the  Reissner-Nordstrom black hole (RNAdS BH) metric by taking a zero of the mass of black hole and cutting of the fifth dimension at infrared (IR) boundary. The deconfinement phase in dual gravity is described by the RNAdS BH metric. There occurs a Hawking-Page transition between these geometries as the confinement-deconfinement phase transition  takes place in the dual field theory \cite{43,44,45,46}.

One of the simplifications in nuclear matter studies using holography is in considering a zero temperature limit. Another one is taking the quark number densities of the medium to be zero and leaving only a dependence on the isospin chemical potential. In the result of these simplifications the isospin medium is obtained for which the background geometry in the dual theory has no modification and metric remains an ordinary AdS space \cite{47,48}. Such a model is useful to separate the effects taking place due to isospin from the ones occurring under the influences of other quantities of the dense medium \cite{1,34,38,39}. One of such effects is the splitting of the mass spectra of the meson states from the same isospin triplet in the medium due to their isospin interaction with the non-zero isospin of the medium. This effect was considered by a number of authors within the holographic QCD \cite{30,34,35,36,47,48,49,50,51}. In \cite{47,48} in the hard-wall model framework it was considered the meson mass splitting effect for the triplets of light mesons $\rho$, $\pi$ and $a_1$ in the dense nuclear matter due to the isospin interaction in connection with the study of the pion condensation in isospin medium, which had been carried out in \cite{34}. In this paper we reconsider this effect because of the recent observation a relation between the definition of the effective mass for a meson in the medium with non-zero isospin and a fixing infrared boundary of hard-wall model. Here we define an effective mass following the standard definition in the field theory  and show that this definition will request fixing the IR boundary differently for the mesons from the same meson isotriplet in the medium. In other words the splitting of meson mass leads to the splitting of the IR boundary of the model, since the mass spectrum in this model is directly defined by the value of the IR boundary. It should be noticed, the IR boundary shift in hard-wall model already is known from the works \cite{37,52}, where the authors deal with the bulk interaction of the fields which contribute to the boundary meson mass.

The paper was organised as follows: In the second section we give a description for the dense and isospin mediums in the hard-wall model. In third section we reproduce the equation of motion for the $\rho $ mesons in isospin medium, determine an effective mass of these mesons by means of holography. We avoid the distinguish of the IR boundary introducing boundary terms. In fourth section similar analysis was made for the $\pi$ and $a_1$ mesons. We discuss relation between mass shift and boundary terms in the last section.
\section{Isospin medium in Hard-wall model}
 Let us present at first a bulk foundation for the boundary isospin medium following  earlier works \cite{42,47,48,53}. The bulk field content for the isospin medium will be a simplified version of the one for the dense nuclear medium utilized in Ref. \cite{47}. Since this medium is described by the $SU(2)$ group, the bulk flavor gauge group should be chosen as $SU\left( 2\right)_{L}\times SU\left( 2\right) _{R}$ which then will broken to $SU(2)_V$. We introduce in the bulk of AdS space two gauge fields $\mathcal{A}^{(L)}$ and $\mathcal{A}^{(R)}$ transforming under the $\left(1,0\right)$  and $\left(0,1\right)$ representations of this flavor symmetry group, respectively. The field strength tensor for these fields is
\begin{equation}
\mathcal{F}_{MN}=\partial _{M}\mathcal{A}_{N}-\partial _{N}\mathcal{A}_{M}-i
\left[ \mathcal{A}_{M},\mathcal{A}_{N}\right],  \label{1}
\end{equation}
with%
\begin{equation}
\mathcal{A}_{M}=\mathcal{A}_{M}^{a}T^{a},\label{2}
\end{equation}
where the flavour index $a$ runs $ a=1,2,3$ and $T^{a}$ are the generators in the spinor representation of the $SU\left( 2\right) _{L}$ or $SU\left( 2\right)_{R}$ group and are expressed in terms of Pauli matrices $\sigma^{a}$ by
\begin{eqnarray}
T^{a}=\frac{1}{2} \sigma^{a}.\label{3}
\end{eqnarray}
 As a gauge condition on $\mathcal{A}_{z}$ we shall choose the axial gauge, $\mathcal{A}_{z}=0$.

The action for the gravity and for the gauge fields will be written as

\begin{equation}
S=\int d^{5}x\sqrt{-G}\left[ \frac{1}{2\kappa ^{2}}\left( \mathcal{R}
-2\Lambda \right) -\frac{1}{4g^{2}}Tr\left( \mathcal{F}_{MN}^{(L)}\mathcal{F}
^{(L)MN}+\mathcal{F}_{MN}^{(R)}\mathcal{F}^{(R)MN}\right) \right] ,
\label{4}
\end{equation}
where $\Lambda =-6/R^{2}$ is the cosmological constant. The AdS/CFT correspondence establishes the following relations between the constants $\kappa ^{2}$, $g^{2}$, the number of colors $N_c$, and the radius $R$ of space-time: $1/g^{2}=N_{c}/\left( 4\pi ^{2}R\right) $ and $1/(2\kappa^{2})=N_{c}^{2}/\left( 8\pi ^{2}R^{3}\right) $.
The gauge fields $\mathcal{A}_{M}^{\left(L\right),\left(R\right)}$ are divided into the background parts $L_{M},R_{M}$ and the fluctuations $l_{M},r_{M}$ of these parts:
\begin{eqnarray}
\mathcal{A}^{\left(L\right)}_{M} &=&L_{M}+l_{M}, \nonumber  \\
\mathcal{A}^{\left(R\right)}_{M} &=&R_{M}+r_{M}.  \label{5}
\end{eqnarray}
In the holographic description of the nuclear matter the bulk flavor group is  $U\left( 2\right)_{L}\times U\left( 2\right) _{R}$ and this matter in the boundary QCD is given by the bulk gauge fields $L_M$, $R_M$, while the fluctuations $l_M$, $r_M$ in the bulk are needed for the description of the vector and axial-vector mesons in the boundary QCD. The homogenous and isotropic matter at the boundary is described by the dual bulk fields  $L_M$ and $R_M$, which are not dependent on space-time coordinates. Moreover, the only time components $L_{0}$ and $R_{0}$ of these fields are taken non-zero, because the only flavor diagonal element of them $\left( a=3\right) $ corresponds to the physical quantity of the boundary matter. From these components we compose the vector and axial vector fields:
\begin{eqnarray}
V_{0}^{3} &=&\frac{1}{2}\left( L_{0}^{3}+R_{0}^{3}\right), \nonumber   \\
A_{0}^{3} &=&\frac{1}{2}\left( L_{0}^{3}-R_{0}^{3}\right).  \label{6}
\end{eqnarray}
The boundary value of $V_{0}^{3}$ maps to the isospin chemical potential of the medium $u$ and $d$ quarks \footnote{In confinement phase these quantities respect to the nucleons.}  \cite{53}.  Following \cite{47, 48} here we shall deal with the case $L_{0}^{a}=R_{0}^{a}$, which means Lagrangian  invariance under changing the left and right flavor groups $SU\left( 2\right) _{L}\leftrightarrow SU\left( 2\right) _{R}$. Obviously, $A_{0}^{3} =0$ for this case. This $L\leftrightarrow R$ invariance in the dual boundary theory means that the medium particles, i.e. the nucleons, are required to be in the parity-even states. As is known, the ground states of the nucleons are parity-even ones, and the first excited state of the nucleon can be  parity-even or parity-odd state. In parity-even state the nucleons have less energy than in parity-odd one \cite{15, 16}.

The field strength tensor for $V^3_{0}$ is a flavor diagonal matrix as well and hence it gets the following form as one for an Abelian field:
\begin{equation}
F_{MN}^{3}=\partial _{M}V_{N}^{3}-\partial _{N}V_{M}^{3}. \label{7}
\end{equation}
In such a way the $SU\left( 2\right) $ symmetry of $V_{M}^a$ part of gauge fields is broken down to two $U\left( 1\right) $ symmetries. Meanwhile, the fluctuations $l_M$, $r_M$  remain to be non-abelian. Then the $V_{M}^3$ part of the action (\ref{2})  gets the form:
\begin{equation}
S=\int d^{5}x\sqrt{-G}\left[ \frac{1}{2\kappa ^{2}}\left( \mathcal{R}
-2\Lambda \right) -\frac{1}{4g^{2}}F_{MN}^{3}F^{3MN} \right]. \label{8}
\end{equation}
The holographic dual of the $A_0^{(u),(d)}=\pm\frac{1}{\sqrt{2}}V_0^3$ field will be the $u$- and $d$-quarks of the boundary medium. Imposing the hard-wall cut-off on the bulk radial coordinate $z$ makes these quarks confined ones in the boundary theory.
It should be noted that the number densities of quarks (isoquarks) thus defined are zero
(see \cite {47}).

In the nuclear matter case the bulk geometry is a thermal charged AdS space (tcAdS) with radius $R$, which is the non-black brane solution of the Einstein-Maxwell system of equations obtained from (\ref{8}) \cite{42}:
\begin{equation}
ds^{2}=\frac{R^{2}}{z^{2}}\left( -f\left(z\right) dt^{2}+\frac{1}{f\left(z\right)}dz^{2}+d\overrightarrow{x}^{2}\right). \label{9}
\end{equation}
Here
\begin{equation}
 f\left(z\right)=1+z^6\sum_{\alpha=u,d}q^2_{\alpha}.\label{10}
\end{equation}
The constants $q_{\alpha}$ is related to the number densities $Q_{\alpha}$ of the medium $u$ and $d$ quarks $q_{\alpha}=\sqrt{2}\kappa Q_{\alpha}/\left(\sqrt{3}gR\right)$. In the isospin medium case the number densities $Q_{\alpha}$ are taken zero and consequently, then the metric (\ref{9}) returns into the metric of ordinary AdS space:
\begin{equation}
ds^{2}=\frac{R^{2}}{z^{2}}\left( - dt^{2}+dz^{2}+d\overrightarrow{x}^{2}\right). \label{11}
\end{equation}
The radial coordinate $z$ ranges in the limited area $0<z\leq z_{IR}$ due to the hard-wall cut-off.

It can be seen from a comparison of the geometries (\ref{9}) and (\ref{11})  in the dual boundary theory that there is an effect of dense medium on the mesons which is provided by the $f\left(z\right)$ function in the bulk metric tensor $G^{MN}$. For instance, if the kinetic energy of $l_M$ or $r_M$ field is $\cal{E}^{\prime}$ in the presence of medium fields $A_{0}^{\left(\alpha\right)}$ and is $\cal{E}$ in the absence of these medium fields, then  $\cal{E}$ and $\cal{E}^{\prime}$ will be related by the relation ${\cal{E}^{\prime}}=G^{00}/G^{\prime 00}{\cal{E}}=f\left(z\right){\cal{E}}$.  Since in the  isospin medium case $Q_{\alpha}=0$ and $f\left(z\right)=1$ there is no modification of the AdS geometry by the medium fields and the effect of the metric (\ref{11}) on the $l_M,$ and $r_M$ fields is same as in free field case.

Solutions to the Maxwell equations obtained from the action (\ref{8}) have the form:
\be
A_0^{(\alpha)}=2\pi^2\mu_{\alpha}-Q_{\alpha}z^2,\quad \alpha=u,\ d. \label{12}
\ee
In the isospin medium case the vector potential $A_0^{(\alpha)}$ is constant and is equal to the quark chemical potential $\mu_{\alpha}$ with a factor $2\pi^2$:
\be
A_0^{(\alpha)}=2\pi^2\mu_{\alpha}. \label{13}
\ee

Here we are going to deal with the medium in the confining phase of the medium $u$ and $d$ quarks, where the fundamental excitations are nucleons but not quarks. So, the solution (\ref{13}) should be expressed in terms of the chemical potentials of the nucleons. Taking into account the quark content of nucleons as in \cite{47} the isospin chemical potentials of nucleons may be defined as a sum of quark chemical potentials $\mu _{P}=2\mu_{u}+\mu _{d}$ and $ \mu _{N}=\mu _{u}+2\mu _{d}$. For isospin matter with two flavors the number densities of the nucleons are zero and the $V_{0}^{3}$ equals
$\sqrt{2}\pi ^{2}\left( \mu _{u}-\mu _{d}\right) $ in the case of deconfinement phase and
\begin{equation}
V_{0}^{3}=\sqrt{2}\pi ^{2}\left( \mu _{P}-\mu _{N}\right) \label{14}
\end{equation}
in the confinement phase one. Thus, $V_0^3$ in the dual boundary theory describes the homogenus and constant isospin background field of medium which is made of isonucleons.

\section{$\rho$ meson mass splitting in the isospin medium}
\subsection{Effective mass definition by use of  holography}
In this subsection applying the AdS/CFT correspondence we shall derive an effective mass formula for the vector meson minimally interacting with the external isospin field (\ref{14}).
It is known that in AdS/QCD models the mesons in dual boundary theory are described by the
fluctuations of the gauge and scalar fields in the bulk. A bulk vector field fluctuation, which corresponds to the boundary vector meson, is composed of the left and right fluctuations $ l_{\mu }^{i}$  and $r_{\mu }^{i}$:
\begin{equation}
v_{\mu }^{i}=\frac{1}{\sqrt{2}}\left( l_{\mu }^{i}+r_{\mu }^{i}\right),\label{15}
\end{equation}
where $\mu=0,1,2,3$ is for the boundary coordinates and $i=1,2,3$ is for the $SU(2)$ index. The fluctuations $l_{M}$ and $r_{M}$ transform under the fundamental representation of $SU\left( 2\right) _{V}$ group ($l_{\mu}=l_{\mu }^{i}\frac{\sigma^i}{2}$ and $r_{\mu}=r_{\mu }^{i}\frac{\sigma^i}{2}$).
As a condition for gauge fixing of the fifth components of the $l_M$ and $r_M$ fields the axial gauge is chosen, which takes values $l_{z}^{i}=r_{z}^{i}=0$. The temporal component of the vector fluctuations will correspond to the fluctuations in the chemical potential (or number density in nuclear matter case) of medium particles. Setting this component zero ($v_{0}^{i}=0$) we shall consider a medium with a constant isospin chemical potential. We suppose the bulk $v_{\mu}$ field does not depend on spatial coordinates $x_{m}$.
The action for the total vector field $\mathcal{V}_M= V_M+v_M=\frac{1}{\sqrt{2}}\left(L_M+R_M+l_{M}+r_{M}\right)$ has the form \cite{47,48}:
\begin{eqnarray}
S_{\mathcal{V}}= -\frac{1}{4g^{2}}\int d^{5}x\sqrt{-G}\ Tr \left( \mathcal{F}_{MN}^{(\mathcal{V})\ast}\ \mathcal{F}^{(\mathcal{V})MN}\right),\label{16}
\end{eqnarray}
where $\mathcal{F}_{MN}^{\mathcal{V}}=\partial_M\mathcal{V}_N
-\partial_N\mathcal{V}_M-i\left[\mathcal{V}_M,\mathcal{V}_N\right]$ is the field strength tensor. After taking the trace in (\ref{16}) the action $S_{\mathcal{V}}$ gets the form:
\begin{eqnarray}
S_{\mathcal{V}}= -\frac{1}{4g^{2}}\int d^{5}x\sqrt{-G}\left\{
\sum\limits_{i=1}^{3}G^{\mu \nu }\left[ G^{zz}\partial _{z}v_{\mu
}^{i\ast}\partial _{z}v_{\nu }^{i}+G^{mn}\partial _{\mu }v_{m}^{i\ast}\partial _{\nu
}v_{n}^{i}\right] \right. \nonumber \\
\left. +G^{00}G^{mn}V_{0}^{3}\left[ V_{0}^{3}\left(
v_{m}^{1\ast}v_{n}^{1}+v_{m}^{2\ast}v_{n}^{2}\right) +2\left( v_{m}^{1\ast}\partial
_{0}v_{n}^{2}-v_{m}^{2\ast}\partial _{0}v_{n}^{1}\right) \right] \right\}.\label{17}
\end{eqnarray}
Here the $m,n=1,2,3$ indices denote the boundary spatial coordinates. The terms in the second square bracket in (\ref{17}) describe the interaction of vector field fluctuations with the background field $V_0^3 $ . In the dual boundary theory this interaction corresponds to the interaction of vector meson with the isospin medium. Thus, the vector meson in isospin medium in a holographic picture is described by the interacting gauge field theory  in the bulk of AdS space.

The complex vector $\rho_m$ field is composed from the vector field components $v_{m}^{i}$  as following:
\begin{eqnarray}
\rho _{m}^{0}&=&v_{m}^{3},\nonumber \\
\rho _{m}^{\pm }&=&\frac{1}{\sqrt{2}}\left( v_{m}^{1}\pm iv_{m}^{2}\right). \label{18}
\end{eqnarray}
As will be seen below the boundary values of the $\rho _{m}^{a}$ bulk fields correspond to the neutral and charged vector mesons respectively.
For the free bulk gauge field $\rho _{m}^{a}$ we can write a Fourier transformation:
\begin{eqnarray}
\rho _{m}^{0}\left( t,z\right) &=&\int \frac{d\omega _{0}}{2\pi }e^{-i\omega
_{0}t}\rho _{m}^{0}\left(\omega _{0}, z \right), \nonumber \\
\rho _{m}^{\pm }\left( t,z\right) &=&\int \frac{d\omega _{\pm }}{2\pi }%
e^{-i\omega _{\pm }t}\rho _{m}^{\pm }\left( \omega _{\pm },z\right),\label{19}
\end{eqnarray}
where the kinetic energies $\omega_a$ of the $\rho^a$ fields
\be
\pa_{0}\pa^{0}\r^a_{n}=\o_a^2\r^a_{n} \quad \left(a=0,\pm \right) \label{20}
\ee
in the dual boundary theory correspond to the masses of the $\rho$ mesons. Equation (\ref{20}) is the one defining a mass in the free field theory. In interacting gauge field theory, where the interaction is included by minimal coupling, the mass of interacting field is defined by means of covariant derivatives instead of ordinary ones in (\ref{20}). Also, as was mentioned above the background gauge field $V_0^3$ does not change the background geometry (\ref{11}). So, the interaction with the $V_0^3$ field still was not included into Eqs. (\ref{19})-(\ref{20}) and the kinetic energies $\omega_{\pm}$ in these formulas are the ones for the free $\rho^a$ field. As is well known, the vacuum masses of all $\rho$ mesons (charged and neutral ones) are equal. This degeneracy takes place for the masses $\o_a$ of the bulk $\rho$ fields in the isospin background $V_0^3$ as well, since there is no metric changes by this background:
 \be
\o_{+}=\o_{-}=\o_{0}. \label{21}
\ee
From the action (\ref{17}) written in terms of $\rho^a$ fields after taking into account (\ref{21}) we will obtain following the equations of motion:
\begin{eqnarray}
&&\partial _{z}\left( \sqrt{-G}G^{zz}G^{mn}\partial _{z}\rho _{n}^{0}\right)
-\omega _{0}^{2}\sqrt{-G}G^{00}G^{mn}\rho _{n}^{0}=0,\nonumber \\
&&\partial _{z}\left( \sqrt{-G}G^{zz}G^{mn}\partial _{z}\rho _{n}^{\pm
}\right) -\left( \omega _{\pm }^{2}+\left( V_{0}^{3}\right) ^{2}\mp 2\omega
_{\pm }V_{0}^{3}\right) \sqrt{-G}G^{00}G^{mn}\rho _{n}^{\pm }=0. \label{22}
\end{eqnarray}
Explicitly, Eq. (\ref{22}) will be written as below:
\begin{eqnarray}
&&\partial _{z}^{2}\rho^{0}_n-\frac{1}{z}\partial _{z}\rho^{0}_n+\omega_{0}^{2}\rho^{0}_n=0, \nonumber \\
&&\partial _{z}^{2}\rho^{\pm }_n-\frac{1}{z}\partial _{z}\rho^{\pm}_n+\left( \omega _{\pm }\mp \sqrt{2}\pi ^{2}\left( \mu _{P}-\mu _{N}\right)
\right) ^{2}\rho^{\pm }_n=0,\label{23}
\end{eqnarray}
which have same form as the ones obtained in \cite{34} for the pions.
These equations of motion were obtained in  \cite{48} introducing $v^i$ fields interacting with the $V_0^3$ field minimally. Lagrangian for this system was constructed by means of covariant derivative $\mathcal{D}_M$ as following:
\begin{equation}
L_{v_M^i+V_0^3}= -\frac{1}{4g^{2}} \left( \mathcal{F}_{MN}^{ i\ast}\ \mathcal{F}^{i MN}\right).\label{24}
\end{equation}
Here $\mathcal{F}_{MN}^{i}$ is defined as $ \mathcal{F}_{MN}^{i}=\mathcal{D}_M v_N^i-\mathcal{D}_N v_M^i$ and the gauge covariant derivative $\mathcal{D}_M$ is $\mathcal{D}_M v_N^i=\partial_M v_N^i+i\left(T^i\right)_{kj}V_M^k v_N^j
=\partial_M v_N^i+\varepsilon^{i3j}V_M^3 v_N^j$. The $\left(T^i\right)_{kj}=-i\varepsilon^{ikj}$
are the generators of isospin group in the adjoint representation. Taking into account the  transverseness of $v^{iM}$ field $\mathcal{D}_M v^{iM}=0$ in (\ref{24}) the action for this Lagrangian will be reduced to the following form:
\begin{equation}
S_{v^i+V_0^3}=-\frac{1}{4g^{2}}\int d^{5}x\sqrt{-G}
 \left(\mathcal{D}_M v^{i\ N}\right)^{\ast} \left(\mathcal{D}^{ M}v^i_N\right).\label{25}
\end{equation}
This action in terms of the $\rho$ fields will split into three independent terms for each $\rho^a$ component:
\begin{equation}
S_{\rho_n^a+V_0^3}=-\frac{1}{4g^{2}}\int d^{5}x\sqrt{-G}\sum_{a=0,\pm}
 \left(D^{(a)}_M\rho^{a\ n}\right)^{\ast} \left(D^{{(a)} M}\rho^a_n\right).\label{26}
\end{equation}
Here the covariant derivative $D^{(a)}_M$ for the $\rho^a$ field has components
$D^{(a)}_{\mu}=\partial_{\mu}+ie^{(a)}V_0^3$, $D^{(a)}_z=\partial_z$ with $e^{(a)}=\pm1,0$. 
In the action (\ref{26}) as in Eq. (\ref{18}), the terms of cubic and forth order of the fluctuations were neglected. The equation of motion obtained from the action (\ref{26}) has the form:
\begin{equation}
D_M^{(a)}\left(\sqrt{-G}G^{MM^{\prime}}G^{mn}D^{(a)}_{M^{\prime}}\rho^a_n\right)=0.\label{27}
\end{equation}
This is a five-dimensional D'Alembert equation in the curved space-time (\ref{11}) for the vector field having zero a fifth component $\left(\rho_5=0\right)$ and zero five-dimensional mass $\left(m_5=0\right)$. Boundary terms which arise on obtaining (\ref{27}) lead to the boundary conditions $\partial_{z}\rho^{a}_{n}|_{z_{IR}}=0$ and $\rho^{a}_{n}\left( z\right)|_{\varepsilon}=0$, which are the same as the ones in the free field case. The explicit form of Eq. (\ref{27}) is the Eq. (\ref{22}).

It is useful to rewrite Eq. (\ref{27}) as a sum of its four-dimensional part and the $z$ coordinate part
\begin{equation}
\frac{1}{\sqrt{-G}}\partial_z\left(\sqrt{-G}G^{zz^{\prime}}G^{mn}\partial_{z^{\prime}}\rho^a_{n}\right)
+\frac{1}{\sqrt{-G}}D_{\m}^{(a)}\left(\sqrt{-G}G^{\m{\m}^{\prime}}G^{mn}D^{(a)}_{{\m}^{\prime}}\rho^a_{ n}\right)=0,\label{28}
\end{equation}
where we divided it by $\sqrt{-G}$. Now, let us write the D'Alembert equation in a four-dimensional curved space-time for the massive vector field $\rho^a_{n}$ minimally coupled with the constant external background gauge field $V_{\m}^a$:
\begin{equation}
\frac{1}{\sqrt{-G}}D_{\m}^{(a)}\left(\sqrt{-G}G^{\m{\m}^{\prime}}G^{mn}D^{(a)}_{{\m}^{\prime}}
\rho^a_{n}\right)-\left(m^{\ast}_a\right)^2G^{mn}\rho^a_{n}=0,\label{29}
\end{equation}
where the $m^{\ast}_a$ is the mass of the $\rho^a_{n}$ field. In our case the four dimensional space-time is the UV boundary of the AdS space (\ref{11}) and the massive vector field is the boundary value of the $\rho$ field. In the AdS/CFT correspondence here the UV values of the $\rho^a$ fields correspond to the $\rho$ mesons defined on this boundary and the equations of motion for these fields at the UV boundary correspond to the equations of motion for these mesons. Hence, we have to correlate Eq. (\ref{28}) equations at UV limit to the ones in (\ref{29}). From this correspondence it is seen that the eigenvalues of the
 $-\frac{1}{\sqrt{-G}G^{mn}}\partial_z\left(\sqrt{-G}G^{zz^{\prime}}G^{mn}\partial_{z^{\prime}}\right)$
operator correspond to the squared mass $\left(m^{\ast}_a\right)^2$ of the $\rho^a$ vector fields in (\ref{29}). So, we may accept the equality of the eigenvalues of this operator to the squared masses of boundary meson fields:
\begin{equation}
-\frac{1}{\sqrt{-G}G^{mn}}\partial_z\left(\sqrt{-G}G^{zz^{\prime}}G^{mn}\partial_{z^{\prime}}
\rho^a_n\right)=\left(m^{\ast}_a\right)^2\rho^a_n.\label{30}
\end{equation}
Eq. (\ref{30}) is the determining formula of effective mass for the vector
field in isospin medium by means of derivative operator over the extra dimension.

In the AdS space geometry (\ref{11}) the equations of motion (\ref{28})  have got the form:
\begin{equation}
\partial_z\left(\frac{1}{z}\partial_z\rho^a_n\right)+\frac{1}{z}D_{\mu}^{(a)}D^{(a)\mu}\rho^a_ n=0,\label{31}
\end{equation}
which is the same as (\ref{23}). The operator $ D_{\mu}^{(a)}D^{(a)\mu}$ has the following eigenvalues
\bea
D_{\m}^{(0)}D^{(0)\m}\r^{0}_n&=&\omega_{0}^{2}\rho^{0}=\left(m^{\ast}_0\right)^2\rho^0_n, \nn
D_{\m}^{(\pm)}D^{(\pm)\m}\r^{\pm}_n&=&\left( \omega _{\pm }\mp \sqrt{2}\pi ^{2}\left( \mu _{P}-\mu _{N}\right)
\right) ^{2}\rho^{\pm }_n=\left(m^{\ast}_{\pm}\right)^2\rho^{\pm }_n.
\label{32}
\eea
In four dimensional space-time, where Eq. (\ref{29}) was written, the eigenvalues (\ref{32}) determine the effective masses of the massive $\rho^a$ fields:
\be
D_{\m}^{(a)}D^{(a)\m}\r_{n}^a=\left(m^{\ast}_a\right)^2\r_{n}^a,\label{33}
\ee
according to the effective mass definition of vector field in the constant background field in four dimensional gauge field theory. The two definitions of the effective mass (\ref{30}) and (\ref{33}) give same result and thus, the masses $m^{\ast}_a$ are the masses of the $\rho^a$ mesons in the isospin medium.

\subsection{Effective mass as an eigenvalue of profile function}
Another way of determining the effective mass of mesons is using profile function solution. From two-point correlation function study it is well known that the eigenvalues in profile function is the mass spectrum of the vector meson in the vacuum case \cite{54}. It is natural to extend this study to the constant isospin medium case. To this end let us write a solution to Eq. (\ref{23}) for the $\rho^{s}_a $ mode in the Kaluza-Klein decomposition $\rho^a\left(p^0,z\right)=\sum_s r^s_a\left(\textbf{m}^{s}_a\right)\rho^s_a\left(\textbf{m}^{s}_a,  z\right)$:
\be
\rho^{s}_a\left(\textbf{m}^{s}_a,  z\right)=\textbf{m}^{s}_az\left[Y_0\left(\textbf{m}^{s}_az_{IR}\right)J_1\left(\textbf{m}^{s}_a z\right)-J_0\left(\textbf{m}^{s}_a z\right)Y_1\left(\textbf{m}^{s}_az_{IR}\right)\right].\label{34}
\ee
Here $\textbf{m}^{s}_a$ denote a square root of bracket in Eq. (\ref{23}) for the $s$ mode:
\bea
\textbf{m}^{s}_0&=&\o_0^s \nonumber\\
\textbf{m}^{s}_{\pm}&=&\o_0^s\mp V_0^3 \label{35}
\eea
and the $\o_a^s$ is a mass of the free $s$ mode.

This solution is the same as the one found in \cite{54} for the free vector field $V_{\mu}$, except for the constant eigenvalues $\textbf{m}^{s}_a$ in arguments of Bessel function, which were denoted by $M_n$ in the free vector field case. Of course,  the distinction between the $M_n$  and $\textbf{m}^{s}_a$ constants will not change the procedure of  finding of the poles by means of the two-point function
\be
\int d^4x e^{ipx}\langle J_{\mu}^a\left(x\right)J_{\nu}^b\left(0\right)\rangle=\delta^{ab}\left(\eta_{\mu\nu}-
\frac{p_{\mu}p_{\nu}}{p^2}\right)\Sigma\left(p^2\right).\label{36}
\ee
Here $\Sigma\left(p^2\right)$ was defined by means of the profile function  $V\left(p,z\right)$ of the vector field as follows
\be
\Sigma\left(p^2\right)=-\frac{1}{g_5^2}\left(\frac{1}{z V\left(p, z\right)}\frac{\partial V \left(p, z\right)}{\partial z}\right)_{z=0}.\label{37}
\ee
In free field case $ V\left(p, z\right)$ has a form of $\rho_s^a\left(\o, V_0^3,  z\right)$ at $V_0^3=0$ in (\ref{34}) and for this solution $\Sigma\left(p^2\right)$ gets the form \cite{54}:
\be
\Sigma\left(p^2\right)=\frac{\pi p^2}{2g_5^2}\left[Y_0\left(p z\right)-J_0\left(p z\right)\frac{Y_0\left(p z_{IR}\right)}{J_0\left(p^2 z_{IR}\right)}\right].\label{38}
\ee
Applying a Kneser-Sommerfeld expansion to Eq. (\ref{38}) in \cite{54} an explicit pole structure of the $\Sigma\left(p^2\right)$  was found:
\be
\Sigma\left(p^2\right)=\frac{2p^2}{g_5^2 z_{IR}^2}\sum_{n=1}^{\infty}\frac{1}{\left[J_1\left(\gamma_0^s\right)\right]^2\left[p^2 -\left(M_n\right)^2\right]}, \label{39}
\ee
where $\gamma_0^s$ are zeros of the Bessel function $J_0\left(z\right)$. For the solution (\ref{34}) the $\Sigma\left(p^2\right)$ in (\ref{39}) will have same expression with the replacement of $M_n$ by $\textbf{m}^{s}_a$ in it and we shall get poles at $p_0^2=\left(\textbf{m}^{s}_a\right)^2$. Since the poles of two-point function are masses of vector mesons in dual theory the poles $\textbf{m}^{s}_a$ will be effective masses of these mesons in isospin medium. The constants $\textbf{m}^{s}_a$ are the eigenvalues of the both $D_{\mu}^{(a)}D^{(a)\mu}$ and $-z\partial_z\left(\frac{1}{z}\partial_z\rho^a_n\right)$ operators for the $s$ mode according to (\ref{33}) and (\ref{31}).

Thus, two ways of determination of the effective masses give the same result and the effective mass of the dual $\rho$ mesons in the isospin medium are the following ones:
\bea
&m^{\ast}_{\pm}&=\mid \o_{\pm} \mp V^3_{0}\mid, \nn
&m^{\ast}_{0}&=\o_{0}.\label{40}
\eea
Although the vacuum masses $\o_a$ of $\rho$ mesons are equal, their in-medium masses
$m^{\ast}_a$ are different. The splitting in the mass spectrum of $\rho$ mesons occur due to interaction with the isospin medium. After taking into account in (\ref{40})  the equality of vacuum mass $\left(\o_{\pm}=\o_{0}\right)$, the splitting formula
will look like:
\be
m^{\ast}_{\pm}=\mid m^{\ast}_{0} \mp V^3_{0}\mid=\mid m^{\ast}_{0}\mp \sqrt{2}\pi
^{2}\left( \mu_{P}-\mu _{N}\right)\mid .\label{41}
\ee
The mass splitting formula (\ref{41}) shows a decreasing of the mass for the positively charged meson and increasing it for the negatively charged one when the positive difference of chemical potentials $\m_P-\m_N$  grows, and vise versa for the negative $\m_P-\m_N$. The isospin medium eliminates the isospin degeneration of mass and degeneration occurs when the chemical potentials of the medium protons and neutrons are equal.

The splitting of meson mass in the holographic approach was studied
in \cite{34,35,47,48}, but the sign of splitting obtained (\ref{41}) is different from the one obtained in \cite{34,47,48}
\bea
&m^{\prime \ast}_{0}&=\o_{0}\nn
&m^{\prime \ast}_{\pm}&=\o_{\pm} \pm V^3_{0}=m^{\prime\ast}_{0}\pm \sqrt{2}\pi ^{2}\left( \mu_{P}-\mu _{N}\right) \label{42}
\eea
and it agrees with the one obtained in \cite{35,55} if we take the value of $V_0^3$  in (\ref{41}) equal to the $\mu_I$ introduced in \cite{35}. It is obvious that the effective mass formula (\ref{42}) does not obey the holographic definition (\ref{30}). This distinction between Eqs. (\ref{41}) and (\ref{42}) is connected with the definition of the effective mass and will be clarified in the next subsection. The sign of the splitting has importance for meson condensation \cite{55}. According to (\ref{41}) the positively charged meson becomes massless at a positive value of the isospin medium, i.e. it condenses at $V_0^3=\o_{+}$, while a condensation of the negatively charged meson takes place at $\o_{-}=-V_0^3$.

\subsection{Mass spectrum and boundary terms}
Applying the Neumann boundary condition at IR boundary $\partial_z\rho^a(z)\mid_{z_{IR}}=0$ on solution (\ref{34}) yields the following formula  (\cite{52}) for the mass spectra $m^{s}_{a}$
\begin{eqnarray}
\textbf{m}^{s}_0&\simeq&\left( s-\frac{1}{4}\right) \frac{\pi}{z_{IR}^{0}},\nonumber \\
\textbf{m}^{s}_{\pm }&\simeq&\left( s-\frac{1}{4}\right) \frac{\pi}{z_{IR}^{\pm}}\quad s=1,2,3...
\label{43}
\end{eqnarray}
which has same form as one in the vacuum and corresponds to mass spectra of excited states of the $\rho$ mesons in the medium. Here we have distinguished the value of IR boundary for the differently charged meson, denoting them by the  $z^{0}_{IR}$ and $z_{IR}^{\pm}$, respectively. Otherwise, the equality of the right hand sides of first and second equalities in (\ref{43}),
$\left(\left( s-\frac{1}{4}\right)\pi/{z_{IR}^{\pm}}=\left( s-\frac{1}{4}\right)  \pi/{z_{IR}^{0}}\right)$
would mean the equality of their left hand sides also,
$\left(\omega _{0}^{s}\mp \sqrt{2}\pi^{2}\left( \mu _{P}-\mu_{N}\right)=\omega _{0}^{s}\right)$,
which is nonsense in non-zero medium isospin case. However, when we introduce different $z_{IR}$ for different components we get an interaction of fields defined in different space-times, which is not available as well. In order to solve this dilemma we can add to the action new boundary terms, which will lead to the same Kaluza-Klein spectrum (\ref{43}) for all $\rho^{a}$ components and therefore will ensure fixing $z_{IR}^{\pm}$ at the same value with $z_{IR}^{0}$. Obviously, new boundary terms should have a form not changing the equation of motion (\ref{22}) and (\ref{23}). We introduce such a kind of boundary terms at $z=z_{IR}$ in the following form:
\be
S_{\pm}=\pm V_0^3\int d^4x \sqrt{-G} G^{00} G^{mn} \left(\left(\rho^{\pm}_m\right)^{\ast}\rho^{\pm}_n\right)_{z=z_{IR}} .\label{44}
\ee
In terms of the $v_{\mu}$ fields, $S_+$ and $S_{-}$ has same expression with opposite sign:
\be
S_{\pm}=\pm  \frac{1}{2}V_0^3\int d^4x \sqrt{-G} G^{00} G^{mn}\left(v^{1\ast}_m v^{1}_n+v^{2\ast}_m v^{2}_n\right)_{z=z_{IR}}.\label{45}
\ee
 Then the total action for the $\rho^a$ fields will have the form
\begin{equation}
S_{\rho+V^3_0}=\int d^{4}x \int_0^{z_{IR}}dz\sqrt{-G}\ \sum_a \mathcal{L}^{(a)}+S_{+}+S_{-}, \label{46}
\end{equation}
where $\mathcal{L}^{(a)}$ is the Lagrangian term producing the equation of motion (\ref{22}) for the  $a$ isocomponent:
\begin{eqnarray}
&&\mathcal{L}^{(0)}=G^{zz}G^{mn}\left(\partial _{z}\left(\rho _{m}^{0}\right)^{\ast}\right)\left(\partial _{z}\rho _{n}^{0}\right)
+\omega _{0}^{2}G^{00}G^{mn}\left(\rho _{m}^{0}\right)^{\ast}\rho _{n}^{0},\nonumber \\
&&\mathcal{L}^{(\pm)}=G^{zz}G^{mn}\left(\partial _{z}\left(\rho _{m}^{\pm}\right)^{\ast}\right)\left(\partial _{z}\rho _{n}^{\pm}\right) +\left( \omega _{\pm }^{2}+\left( V_{0}^{3}\right) ^{2}\mp 2\omega
_{\pm }V_{0}^{3}\right) G^{00}G^{mn}\left(\rho _{m}^{\pm}\right)^{\ast}\rho _{n}^{\pm }. \label{47}
\end{eqnarray}

When we derive the equations of motion (\ref{22}) from the (\ref{47}) the boundary terms
arising on integration by parts have the same form $\left(\rho^a\right)^{\ast}\partial_z \rho^a\mid^{z_{IR}}_{z_{UV}}$ for the different $\rho^a$ components. These boundary terms will be summed with the $S_{\pm}$ boundary terms and the total boundary condition imposed on the $\rho^a$ solutions at the IR boundary ultimately will be written as below:
 \begin{equation}\label{48}
\left( \partial_z \rho^{0}\left(z\right)\right)\mid_{z_{IR}}=0,\quad
\left( \partial_z \rho^{+}\left(z\right)+V_0^3\rho^{+}\left(z\right)\right)\mid_{z_{IR}}=0,\quad
\left( \partial_z \rho^{-}\left(z\right)-V_0^3\rho^{-}\left(z\right)\right)\mid_{z_{IR}}=0.
 \end{equation}
All three boundary conditions gives same quantisation formula - the same spectrum (\ref{43}) but for the vacuum masses $\o_0^s$ of in-medium $\rho $ mesons:
\begin{eqnarray}
\textbf{m}^{s}_{0}=\o_0^s&\simeq&\left( s-\frac{1}{4}\right) \frac{\pi}{z_{IR}},\nonumber \\
\textbf{m}^{s}_{\pm}\pm V_0^3=\o^{s}_{\pm}\mp V_0^3 \pm V_0^3=\o_0^s&\simeq&\left( s-\frac{1}{4}\right) \frac{\pi}{z_{IR}}\quad s=1,2,3...
\label{49}
\end{eqnarray}
and the same $z_{IR}$ will be chosen in all three quantization formulas (\ref{49}). Thus, we conclude that only the vacuum mass part $\o_0^s$ of the effective mass of meson is subject to quantization.

Now we can explain why there are two ways of the definition of the meson mass splitting (Eqs. (\ref{41}) and (\ref{42})). The splitting formula (\ref{42}) was obtained in the hard-wall model which has one IR boundary for the all isocomponents of the vector field. In this case all three equations in (\ref{23}) were solved under the same boundary
condition at IR and with the same value of $z_{IR}$. This dictates that all coefficients in front of last terms in the equations (\ref{23}) should be the same. This leads to the splitting formula (\ref{42}). In fact, if we put back (\ref{42}) into (\ref{23}) we shall get the same expression for these coefficients, which equals to $\o_0$, i.e. we have same three equations for $\rho^{\pm}$ and $\rho^{0}$. Obviously, in this case the all $\textbf{m}^{s}_a$ coefficients in Bessel function solutions are the same $\textbf{m}^{s}_a=\omega_0$ and the necessity  does not arise to distinguish $z_{IR}$ in the mass spectrum formula (\ref{43}). In (\ref{44}) we obtained this spectrum formula introducing boundary terms, but the effective mass in our case obeys the defining relations (\ref{30}) and (\ref{33}).

The first state ($s=1$) of charged vector field tower (\ref{40}) is the ground state of the $\rho^{\pm} $ meson, which has a mass
\begin{equation}
m_{\rho^{\pm}}\approx \frac{3}{4}\pi \frac{1}{z_{IR}^{0}}\mp\sqrt{2}\pi^{2}\left( \mu _{P}-\mu_{N}\right) \approx 2.4z_{IR}^{-1}\mp\sqrt{2}\pi^{2}\left( \mu _{P}-\mu_{N}\right)
\label{50}
\end{equation}
and for the neutral $\rho$ meson we have the value of mass in the vacuum \cite{52}:
\begin{equation}
m_{\rho^{0}}\approx \frac{2.4}{z_{IR}^{0}}\label{51}
\end{equation}
Finally, the normalised solution (\ref{34}) for the $\rho^a$ meson attains the same form as the one in vacuum case \cite{52}:
\begin{equation}
\rho_a^{s}=\frac{zJ_1\left(\textbf{m}^{s}_a z\right)}{\sqrt{\int_0^{z_{IR}} dzz\left[J_1\left(\textbf{m}^{s}_a z\right)\right]^2}}.\label{52}
\end{equation}

\section{Mass splitting formula for the $a_{1}$ and $\pi$ mesons}
In this section we shall define the effective mass for the axial-vector and  pion fields using holographic correspondence as was done for the $\rho $ mesons. Let us briefly present here the action and the equations of motion for the axial-vector and pion fields, all details of which can be found in \cite{47,48}.

The non-zero fluctuations of the axial-vector field, which is defined as $a_{\mu}^i=
\frac{1}{\sqrt{2}}\left(l_{\mu}^i-r_{\mu}^i\right)$, were introduced into the model in order to describe the axial-vector mesons in the boundary QCD. In addition, a complex scalar field $\Phi$  is introduced, which performs the chiral symmetry breaking $SU\left( 2\right)_{L}\times
SU\left( 2\right)_{R}\rightarrow SU\left( 2\right)_{V}$ of the model. The action for the
$\Phi$ field has the form:
\be
S_{\phi}=-\int d^{5}x\ \sqrt{-G}\ Tr\left[ \left|D \Phi\right|^2
+m_5^2\left|\Phi\right|^2\right].\label{53}
\ee
Here $D_M \Phi = \partial_M \Phi -iL_M\Phi+i\Phi R_M$ and $m_5^2=-3$. The $\Phi$ field is
written in the form: $\Phi=\phi\ exp \left[i\sqrt{2}\pi^aT^a\right]$. The fluctuations $\pi^i$ describe the pions in the dual theory.
In isospin medium case, as was mentioned, the background geometry is given (\ref{11}) and thus the solution for the $\phi$ field has  a simple form of one in vacuum case:
\begin{equation}
\phi\left(z\right)=\frac{1}{2}m_{q}z+\frac{1}{2}\sigma z^3,\label{54}
\end{equation}
where $m_q$ is the mass of light quarks and the $\sigma$ is the value of the condensate.
The action (\ref{53}) contains an interaction of $\Phi$ field with the axial-vector field
$a_{\mu}^i$ and the total action of these fields will be the sum of $S_{\phi}$ and the action $S_{a}$ for the fluctuations $a_{\mu}^i$. The action $S_{a}$, which describes the axial-vector field interacting with the gauge field $V_0^3$ is obtained from the action (\ref{4}) and has the form:
\begin{eqnarray}
S_{a}= -\frac{1}{4g^{2}}\int d^{5}x\sqrt{-G} \left( \mathcal{F}_{MN}^{a\ \ast}\ \mathcal{F}^{a \ MN}\right),\label{55}
\end{eqnarray}
where $ \mathcal{F}_{MN}^{i}=\mathcal{D}_M a_N^i-\mathcal{D}_N a_M^i$ and the gauge covariant derivative $\mathcal{D}_M$ is $\mathcal{D}_M a_N^i=\partial_M a_N^i+i\left(T^i\right)_{kj}V_M^k a_N^j
=\partial_M a_N^i+\varepsilon^{i3j}V_M^3 a_N^j$ was defined as one for the vector field fluctuations $v^i$.
Since in the dual QCD theory the axial-vector $a_1$ mesons are described by the transverse
field we divide the bulk axial-vector field $a_{\mu}^i$ into the transverse $\bar{a}_{\mu}^i$ and the longitudinal $\chi$ parts:
 $a_{\mu}^i=\bar{a}_{\mu}^i+ \partial_{\mu}\chi^i \quad \left(\partial^{\mu}\bar{a}_{\mu}=0\right)$.
 The longitudinal parts of the left and right fluctuations $l_{\mu}^i$ and $r_{\mu}^i$ are equal to  $+\frac{1}{2}\partial_{\mu}\chi^i$ and $-\frac{1}{2}\partial_{\mu}\chi^i$ respectively and the vector field fluctuations $v^i_{\mu}$ turns out transverse. In an equivalent way an action for the transverse axial-vector fluctuations $\bar{a}_{\mu}^i$ can be obtained from the action (\ref{4}) and has the form \cite{47}:
\begin{eqnarray}
S_{\bar{a}} = -  \frac{1}{4 g^2} \int d^5 x \sqrt{- G}\ G^{m n} \left\{ \sum_{i=1}^{3}
\left[ G^{zz} \partial_z\bar{a}^{i\ast}_{m} \partial_z \bar{a}^i_{n}
 + G^{\mu \nu}  \partial_{\mu} \bar{a}^{i\ast}_{m}  \partial_{\nu} \bar{a}^{i}_{n}  \right] \right. \nonumber \\
 \left. + G^{00} V_0^3 \left[V_0^3 \left( \bar{a}^{1\ast}_{m} \bar{a}^1_{n} +
 \bar{a}^{2\ast}_{m} \bar{a}^2_{n}  \right)
+ 2  G^{00}  \left( \bar{a}^{1\ast}_m \partial_0 \bar{a}^2_{n}-\bar{a}^{2\ast}_m
\partial_0 \bar{a}^1_n  \right)\right]
 \right\}.\label{56}
\end{eqnarray}
The two actions (\ref{55}) and (\ref{56}) lead to the same equations of motion for the $\bar{a}$ field, as in the vector field case.
It is recommended to incorporate the action for the longitudinal $\chi$ field with the action for the pseudoscalar field  $\pi$ derived from (\ref{53}). In summary the action of $\chi$ and $\pi$ fields has the following explicit form \cite{47}:
\begin{eqnarray}
&S_{\pi+\chi} = -  \frac{1}{4 g^2} \int d^5 x \sqrt{- G}\left\{  \sum_{i=1}^{3}
\left[ G^{zz}G^{\mu\nu}\partial_z\partial_{\mu}\chi^{i\ast}\partial_z\partial_{\nu}\chi^i\right]
+4g^2\phi^2\left[ \sum_{i=1}^{3}  \left(G^{\mu\nu}\partial_{\mu}\chi^{i\ast}\partial_{\nu}
\chi^i\right. \right.\right.&\nonumber\\
&\left.\left.\left. + G^{z z} \partial_z \pi^{i\ast}  \partial_z \pi^i+G^{\mu \nu}\partial_{\mu}
 \pi^{i\ast}  \partial_{\nu}\pi^i -2G^{\mu \nu}\partial_{\mu} \chi^{i\ast} \partial_{\nu}\pi^i  + G^{mn}\bar{a}^{i\ast}_{m}\bar{a}^i_{n}\right)+ G^{00}V^3_0  \left[ 2 \left( \pi^{1\ast}  \partial_0 \pi^2 - \pi^{2\ast}  \partial_0 \pi^1 \right) \right. \right. \right.&\nonumber \\
&\left.\left.\left. +V^3_0 \left( \left|\pi^1\right|^2 + \left|\pi^2\right|^2\right)-
2\left(\pi^{1\ast}\partial_0\chi^2-\pi^{2\ast}\partial_0\chi^1\right)\right]\right]\right\}.
\label{57}&
\end{eqnarray}
This action includes the interaction term of the $\bar{a}^i$ field with the $\phi$ field, which is coming from the action (\ref{53}) and obviously, this term will be present in the equation of motion for the $a_1$ field.
\subsection{the $a_1$ mesons}

Let us introduce the neutral $a_{1m}^{0}$ and the
charged $a_{1m}^{\pm}$ components for the axial-vector field $\bar{a}_m$, which will correspond to the respective axial-vector  $a_{1}$ meson in the holography:
\begin{eqnarray}
a_{1m}^{0}&=&\bar{a}^{3}_{m}, \nonumber \\
a_{1m}^{\pm}&=&\frac{1}{\sqrt{2}}\left(\bar{a}^{1}_{m}\pm i\bar{a}^{2}_{m}\right). \label{58}
\end{eqnarray}
Tre Fourier transforms of these components are
\begin{eqnarray}
a_{1m}^{0}\left( t,z\right) &=&\int \frac{d\bar{\omega }_{0}}{2\pi }e^{-i\bar{\omega }_{0}t}a_{1m}^{0}\left(\bar{\omega }_{0}, z \right), \nonumber \\
a_{1m}^{\pm }\left( t,z\right) &=&\int \frac{d\bar{\omega }_{\pm }}{2\pi }%
e^{-i\bar{\omega }_{\pm }t}a_{1m}^{\pm }\left(\bar{\omega }_{\pm },z\right), \label{59}
\end{eqnarray}
where $\bar{\omega}_{0}$ and $\bar{\omega}_{\pm } $ are the masses of free $a_{1}$ fields.
The action for the $a_1^a$ field will be written in the form:
\begin{equation}
S_{a_1^a+V_0^3}=-\frac{1}{4g^{2}}\int d^{5}x\sqrt{-G}\sum_{a=0,\pm}\left\{
 \left(D^{(a)}_M \ a_1^{a\ n}\right)^{\ast} \left(D^{{(a)} M}a^a_{1\ n}\right)+4g^{2}\phi^{2}a_{1}^{a\ n} a_{1\ n}^{a}\right\}, \label{60}
\end{equation}
where the covariant derivative $D^{(a)}_M$ is the same as in the vector field case. the equations of motion for this field written in terms of $D_M^{(a)}$ will be obtained from the action (\ref{60}) and has the form below:
\begin{equation}
D_M^{(a)}\left(\sqrt{-G}G^{MM^{\prime}}G^{mn}D^{(a)}_{M^{\prime}}a_{1n}^a\right)
+4g^{2}\phi^{2}\sqrt{-G}G^{mn}a_{1n}^{a}=0.
\label{61}
\end{equation}
This is a five-dimensional D'Alembert equation  for the axial-vector field interacting with the $\phi(z)$ field. Obviously, the fifth component and five-dimensional mass $M_5$ of this field are taken to be zero. The boundary conditions, which will be imposed on $a_{1n}$
are obtained from the boundary terms arising on deriving Eq. (\ref{61})
and are usual ones: $\partial_{z}a_{1n}^a|_{z_{IR}}=0$ and $a_{1n}^a\left( z\right)|_{\varepsilon}=0$.

It will be useful to rewrite Eq. (\ref{61}) as the sum of its four-dimensional part and the fifth dimension part:
\begin{equation}
\frac{1}{\sqrt{-G}}\partial_z\left(\sqrt{-G}G^{zz^{\prime}}G^{mn}\partial_{z^{\prime}}
a_{1n}^{a}\right)+\frac{1}{\sqrt{-G}}D_{\m}^{(a)}\left(\sqrt{-G}G^{\m{\m}^{\prime}}
G^{mn}D^{(a)}_{{\m}^{\prime}}a_{1n}^{a}\right)+4g^{2}\phi^{2}G^{mn}a_{1n}^{a}=0.
\label{62}
\end{equation}
 On the QCD side there is a massive axial-vector meson, which is described by the $a_{1n}$ field and has a shift in the mass due to the quark condensate background. We may take into account this mass shift by adding to the QCD Lagrangian additional mass term proportional to the value of condensate, which will be denoted by the $4g^{2}\phi_0^{2}$. The constant isospin background of the medium can be taken into account by introducing a constant gauge field $V_0^3$. The interaction with this background will be taken into account by the  minimal interaction. Let us write the D'Alembert equation for such an $a_{1n}$ field in a four-dimensional curved space-time, which is, in fact, the UV boundary of AdS space (\ref{11}):
\be
\frac{1}{\sqrt{-G}}D_{\m}^{(a)}\left(\sqrt{-G}G^{\m{\m}^{\prime}}G^{mn}D^{(a)}_{{\m}^{\prime}}
a^a_{1n}\right)+4g^{2}\phi_0^{2}G^{mn}a_{1n}^{a}-\left(\bar{m}^a\right)^2G^{mn}a^a_{1n}=0.
\label{63}
\ee
Here $\bar{m}^a$ is the mass of the $a^a_{1n}$ field not including the condensate shift and $\phi_0$ is the value of $\phi(z)$ field at some fixed value of $z$. Since the AdS/CFT correspondence of the fields takes place at the UV boundary we take as a fixed value of $z$ the $z_{UV}$, i.e. $\phi_0=\phi\left(z_{UV}\right)$. The condensate interaction term $4g^{2}\phi_0^{2}G^{mn}a_{1n}^{a}$ in (\ref{63}) could be included into the mass term by $\left(m_{a}^2=\left(\bar{m}^a\right)^2-4g^{2}\phi_0^{2}\right)$.
Correlating Eqs. (\ref{62}) to (\ref{63}) we observe the equality of mass term in (\ref{62}) and $z$-dependent part of Eq. (\ref{63})
\begin{equation}
\left(\bar{m}^a\right)^2G^{mn}a^a_{1n}=
-\frac{1}{\sqrt{-G}}\partial_z\left(\sqrt{-G}G^{zz^{\prime}}G^{mn}\partial_{z^{\prime}}
a_{1n}^{a}\right).
\label{64}
\end{equation}
Eq. (\ref{64}) is the formula determining the effective mass of the axial-vector meson in the isospin medium by means of the derivative operator over the extra dimension defined
in dual bulk theory.
The equations of motion for the  $a_{1}^a$ field are obtained from the summary action of (\ref{63}) and (\ref{64}) and given by \cite{47}:
\begin{eqnarray}
&&\partial _{z}\left( \sqrt{-G}G^{zz}G^{mn}\partial _{z}a _{1n}^{0}\right)
-\left(G^{00}\bar{\omega }_{0}^{2}+4g^{2}\phi^{2}\right)\sqrt{-G}G^{mn}a_{1n}^{0}=0, \nonumber \\
&&\partial _{z}\left( \sqrt{-G}G^{zz}G^{mn}\partial _{z}a_{1n}^{\pm}\right)
-\left[\left(\bar{\omega }_{\pm }\mp V_{0}^{3}\right)^{2}G^{00}
+4g^{2}\phi^{2}\right] \sqrt{-G}G^{mn}a_{1n}^{\pm }=0. \label{65}
\end{eqnarray}
The explicit form of Eq. (\ref{65}) in the background geometry (\ref{11}) after the replacement $a_{1}^{a}(z)=za_{1}^{\prime a}(z)$  will get the following form:
 \begin{eqnarray}
&&\partial _{z}^{2}a_{1}^{\prime0}+\frac{1}{z}\partial _{z}a_{1}^{0}+\left[\bar{\omega }_{0}^{2}-g^{2}\left(m_{q}+\sigma z^2\right)^{2}-\frac{1}{z^2}\right]a_{1}^{\prime0}=0, \nonumber \\
&&\partial _{z}^{2}a_{1}^{\prime\pm }+\frac{1}{z}\partial _{z}a_{1}^{\pm}+
\left[\left(\bar{\omega}_{\pm }\mp V^{3}_{0}\right)^{2}-g^{2}\left(m_{q}
+\sigma z^2\right)^{2}-\frac{1}{z^2}\right]a_{1}^{\prime\pm }=0,
\label{66}
\end{eqnarray}
Following earlier work \cite{47,48} we consider at first the case of ignoring the condensate term $4g^2\phi^2$ in the equations of motion (\ref{66}). In this case we have no shift of squared mass and a solution to Eq. (\ref{66}) for the $s$-th mode $a_{1 s}^{a}(z)$ in Kaluza-Klein decomposition $$ a_{1n}^a=\sum_s f_{1n}^{a s}(P^0)a_{1 s}^{a}(P^0, z)$$ after applying UV boundary condition gets a form, which is same as the one for the vector mesons:
 \begin{equation}
a_{1 s}^{a}(z)=c_{1}zJ_{1}\left(\bar{m}^{s}_{a}z\right). \label{67}
\end{equation}
The mass spectra $\bar{m}^{s}_{a}$ of the $s$ states are
\begin{eqnarray}
\bar{m}^{s}_{0}&=\bar{\omega}_{0}^s, \nonumber \\
\bar{m}^{s}_{\pm}&=\mid\bar{\omega}_{\pm}^s\mp V_{0}^{3}\mid
\label{68}
\end{eqnarray}
and $\bar{\omega }_{\pm}^s=\bar{\omega}_{0}^s$ are the vacuum masses of the excited axial-vector mesons. From (\ref{68}) the mass splitting formula for the $a_1^{\pm}$ mesons will be written in the following form:
\be
\bar{m}^{s}_{\pm}=\mid\bar{m}^{s}_{0}\mp V_0^3\mid.
\label{69}
\ee
It is obvious the mass spectrum formula (\ref{43}) is available for the axial-vector field case as well. The distinction in $z_{IR}$ can be compensated by introducing extra boundary terms $S_{\pm}$ similar to the ones for the vector meson case:
\be
S_{\pm}=\pm V_0^3\int d^4x \sqrt{-G} G^{00} G^{mn} \bar{a}^{\pm \ast}_{1m} \bar{a}^{\pm}_{1n}\mid_{z=z_{IR}} .\label{70}
\ee
These boundary terms will shift back the value of infrared boundary $z_{IR}^{\pm}$ to the same value $z_{IR}$ for all $a^a_1$ mesons.
In the case, when $\phi^2\neq0$ Eq. (\ref{60}) can be solved in the UV ($z\rightarrow0$) and in the IR ($z\rightarrow z_{IR}$)  limits. In order to find mass spectrum in this case it is reasonable to apply the IR boundary condition to the asymptotic solution found in the IR limit. For the IR asymptotic solution we shall take $z\rightarrow z_{IR}$ limit from (\ref{60}) and  set the $z=z_{IR}$ in the condensate term. Before performing this approximation let us  compare numerically the last two terms in Eq. (\ref{60}) when $z\rightarrow z_{IR}$.
The approximate values are: $z_{IR}^{-1}\approx0.33$ $\left(GeV\right)$, $z_{IR}\approx3$ $\left(GeV\right)^{-1}$, $z_{IR}^{4}\approx81$
$\left(GeV\right)^{-4}$, $\sigma\approx\left(0.3\right)^3$ $\left(GeV\right)^3$, $g^2=4\pi^2/N_c\approx13.2$. Then
$g^2\sigma^2\left(z_{IR}\right)^4\approx0.06$ $\left(GeV\right)^2$ and $1/\left(z_{IR}\right)^2=0.1$
$\left(GeV\right)^2$. Thus, the $1/z^2$ term contributes twice more than the $g^2\sigma^2 z^4$ term and so we may make an approximation in (\ref{60}) by setting $z=z_{IR}$ only in the condensate term and keeping the term $1/z^2$ variable. At this limit the condensate term in the equations (\ref{60}) becomes constant and the IR asymptotic solution of these equations is expressed in terms of Bessel function $J_1$:
 \be
a_{1 s}^{a}=czJ_1\left(\bar{m}^s_az\right).\label{71}
\ee
Obviously, the UV boundary condition was applied on this solution. The mass spectrum $\bar{m}^s_a$ in (\ref{71}) is expressed in terms of $z_{IR}$ :
\begin{eqnarray}
\bar{m}^{s}_{0}&=&\sqrt{\left(\bar{\omega}_{0}^s\right)^2-g^{2}\left[m_{q}+\sigma
\left(z_{IR}\right)^{2}\right]^{2}}\approx\bar{\omega}_{0}^s-\frac{g^{2}\left[m_{q}+
\sigma\left(z_{IR}\right)^2\right]^{2}}{2\bar{\omega }_{0}^s}, \nonumber \\
\bar{m}_{\pm}^s&=&\sqrt{\left(\bar{\omega }_{\pm }^s\mp V^{3}_{0}\right)^{2}
-g^{2}\left[m_{q}+ \sigma\left(z_{IR}\right)^2\right]^{2}}\approx
\mid\bar{\omega}_{\pm}^s\mp V^{3}_{0}\mid-\frac{g^{2}\left[m_{q}
+\sigma\left(z_{IR}\right)^2\right]^{2}}{2\mid\bar{\omega}_{\pm}^s\mp V^{3}_{0}\mid}.
\label{72}
\end{eqnarray}
As is seen from (\ref{72}) the condensate term contributes to the mass of states with different $s$ differently. The mass splitting formula for the spectra (\ref{72}) will be written as usual:
\be
\bar{m}^{s}_{\pm}\approx \mid \bar{m}^{s}_{0}\mp V_0^3+ \frac{g^{2}\left(m_{q}+\sigma\left(z_{IR}\right)^{2}\right)^2}{2\bar{\omega}_{0}^s} \mid-\frac{g^{2}\left(m_{q}+\sigma \left(z_{IR}\right)^{2}\right)^2}{2\mid \bar{\omega }_{\pm}^s\mp V_{0}^{3}\mid}
\label{73}
\ee
and has the splitting sign as in the case of  absence of the condensate.
In the case taking into account the condensate term the mass spectrum formula (\ref{41}) is valid for the mass shifted by condensate $\bar{m}^{s}_{a}$:
\begin{eqnarray}
\bar{m}^{s}_{0}&\simeq&\left( s-\frac{1}{4}\right) \frac{\pi}{z_{IR}^{0}},\nonumber \\
\bar{m}^{s}_{\pm }&\simeq&\left( s-\frac{1}{4}\right) \frac{\pi}{z_{IR}^{\pm}}\quad s=1,2,3...
\label{74}
\end{eqnarray}
This time a shift in mass is not simply $\mp V_0^3$ as in the previous case and it has a contribution proportional to the condensate value $\phi\left(z_{IR}\right)$. This contribution depends on $s$ and therefore,  the $z_{IR}^{\pm}$ this time depends on $s$ as well. This complicates the introduction of the IR boundary terms compensating the $z_{IR}^{\pm}$ distinguished with the $z_{IR}^{0}$ , which would be unique for all $s$ states. For this case we have to introduce the boundary term for each state $s$, which depends on mass $\bar{\omega }_{\pm}^s$ of this state as well,
\be
S_{\pm}^s=\left(\pm V_0^3+ \frac{g^{2}\left(m_{q}+\sigma\left(z_{IR}\right)^{2}\right)^2}{2\mid \bar{\omega }_{\pm}^s\mp V_{0}^{3}\mid}\right)\int d^4x \sqrt{-G} G^{00} G^{mn} f^{\pm s \ast}_{1 m} f^{\pm s}_{1 n}\left(a_{1 s}^{\pm}\right)^2\mid_{z=z_{IR}} \label{75}
\ee
and total boundary term will be sum of boundary terms of the $s$ modes $$S_{\pm}=\sum_s S_{\pm}^s.$$
We should compensate as well the shift in the mass spectrum of neutral $a_1^0$ meson, which was caused by the condensate term. As is seen from Eq. (\ref{72}) the condensate shift in the $\bar{m}^{s}_{0}$ spectrum is different for the different $s$ states. This means one should distinguish values of $z_{IR}$, these being also $s$ dependent. In order to avoid the distinguished $z_{IR}$ we introduce an additional boundary term for the each Kaluza-Klein $s$  mode and sum these terms:
\be
S_{0}= \frac{1}{2}g^{2}\left(m_{q}+\sigma\left(z_{IR}\right)^{2}\right)^2\int d^4x \sqrt{-G} G^{00} G^{mn}\sum_s \frac{1}{ \bar{\omega }_{0}^s}f^{0 s \ast}_{1 m} f^{0 s}_{1 n}\left(a_{1 s}^{0}\right)^2\mid_{z=z_{IR}} .\label{76}
\ee

 One should notice that in \cite{52}, where the axial-vector meson's mass spectrum including condensate contribution was obtained in the vacuum case, an approximation was performed which shifts the boundary condition at $z_{IR}$ and, thus, the condensate contribution to the mass of meson was taken into account. To do a shift in such a way is equivalent to introducing the boundary term $S_{0}$ in our case\footnote{Although the condensate shift in the squared mass spectrum is the same for all Kaluza-Klein states $s$ the approximation here takes it into account mass spectrum depending on mass $\o^s$ of the state. In the approximation made in \cite{52} the shift in the IR boundary condition is same for all $s$ states.}.
\subsection{$\pi$ mesons in isospin medium}

Following \cite{47}, we introduce the charged $\pi^a$ and $\chi^a$ fields by means of  $\pi^i$ and $\chi^i$ fields, respectively, as below:
\begin{eqnarray}
\pi _{m}^{0}&=&\pi_{m}^{3},\quad \chi _{m}^{0}=\chi_{m}^{3},\nonumber \\
\pi_{m}^{\pm }&=&\frac{1}{\sqrt{2}}\left( \pi_{m}^{1}\pm i\pi_{m}^{2}\right), \quad
\chi_{m}^{\pm }=\frac{1}{\sqrt{2}}\left( \chi_{m}^{1}\pm i\chi_{m}^{2}\right). \label{77}
\end{eqnarray}
In the AdS/CFT correspondence the UV boundary values of the $\pi^a$ fields $(a=0,\pm)$ will be mapped to the pion fields of the dual QCD. Fourier components for the $\pi^a$ and $\chi^a$ fields are written as
\begin{eqnarray}
\pi _{m}^{0}\left( t,z\right) &=&\int \frac{d\widetilde{\omega} _{0}}{2\pi }
e^{-i\widetilde{\omega }_{0}t}\pi_{m}^{0}\left(\widetilde{\omega}_{0}, z \right), \quad
\chi _{m}^{0}\left( t,z\right) =\int \frac{d\widetilde{\omega} _{0}}{2\chi}
e^{-i\widetilde{\omega}_{0}t}\chi_{m}^{0}\left(\widetilde{\omega}_{0}, z \right),\nonumber \\
\pi _{m}^{\pm }\left( t,z\right) &=&\int \frac{d\widetilde{\omega}_{\pm }}{2\pi }%
e^{-i\widetilde{\omega }_{\pm }t}\pi_{m}^{\pm }\left(\widetilde{\omega }_{\pm },z\right), \quad
\chi _{m}^{\pm }\left( t,z\right) =\int \frac{d\widetilde{\omega}_{\pm}}{2\pi }%
e^{-i\widetilde{\omega}_{\pm }t}\chi_{m}^{\pm }\left(\widetilde{\omega}_{\pm },z\right),
\label{78}
\end{eqnarray}
where $\widetilde{\omega}_{a}$ are the masses of $\pi^a$ (and $\chi^a$) fields in the absence
isospin background $V_0^3$, i.e. are the vacuum masses and do not include the
contribution of the condensate. As was in the $\rho^a$ and $a_1^a$ meson cases the vacuum masses $\widetilde{\omega}_a$ of the $\pi^a$ fields do not depend on $\mu_P$ and $\mu_N$ and their equality $\widetilde{\omega }_{\pm}=\widetilde{\omega }_{0}$ is right in isospin medium case as well.
The action (\ref{57}) in terms of $\pi^a$ and $\chi^a$ fields splits into three independent  parts:
\begin{equation}
S_{\pi+\chi}=\int d^{4}x\int_0^{z_{IR}}dz \ \sqrt{-G} \left\{ \widetilde{\mathcal{L}}^{(0)}+ \widetilde{\mathcal{L}}^{(+)}
+\widetilde{\mathcal{L}}^{(-)}\right\}, \label{79}
\end{equation}
where
\begin{eqnarray}
\widetilde{\mathcal{L}}^{(0)}=\frac{1}{4g^2}G^{zz}G^{00}\omega_0^2\left(\partial_z\chi^0\right)^2 +\phi^2 G^{zz}\left(\partial_z\pi^0\right)^2+\phi^2 \omega_0^2 G^{00}\left(\chi^0-\pi^0\right)^2,
 \nonumber \\
\widetilde{\mathcal{L}}^{(\pm)}=\frac{1}{4g^2}G^{zz}G^{00}\omega_{\pm}^2\left(\partial_z
\chi^{\pm}\right)^2+ \phi^2 G^{00}\omega_{\pm}^2\left(\chi^{\pm}\right)^2 + \phi^2 G^{zz}
\left(\partial_z \pi^{\pm}\right)^2  \nonumber \\
 +\phi^2 G^{00}\pi^{\pm}\left(\omega_{\pm}\mp V_0^3 \right)\left[\pi^{\pm}
\left(\omega_{\pm}\mp V_0^3 \right)-2\chi^{\pm}\omega_{\pm} \right]. \label{80}
\end{eqnarray}

Equations of motion for these scalar fields are obtained  in \cite{47} from the action (\ref{79}) and are summarized as follows:
\bea			\label{81}
 \chi^0 - \pi^0  &=& \frac{1}{ 4 g^2  \ph^2 \sqrt{-G} G^{00} } \ \pa_z
\ls \sqrt{-G} G^{zz} G^{00} \pa_z \chi^0 \rs , \nn
\o_0^2 \ls \chi^0  -  \pi^0  \rs &=& - \frac{1}{\ph^2 \sqrt{-G} G^{00} }
\ \pa_z \ls \ph^2 \sqrt{-G} G^{zz} \pa_z  \pi^0 \rs ,
\eea
and
\bea                 \label{82}
\o_{\pm} \chi^{\pm} - \ls \o_{\pm} \mp V_0^3  \rs \pi^{\pm}
&=& \frac{\o_{\pm} }{4 g^2  \ph^2 \sqrt{-G} G^{00}} \pa_z \ls \sqrt{-G} G^{zz} G^{00}
\pa_z \chi^{\pm} \rs , \nn
\left( \o_{\pm} \mp V_0^3 \right) \left[ \o_{\pm}  \chi^{\pm} - \left( \o_{\pm} \mp V_0^3  \right) \pi^{\pm} \right]
&=&   - \frac{1}{\ph^2 \sqrt{-G} G^{00}} \pa_z \left( \ph^2 \sqrt{-G} G^{zz} \pa_z
\pi^{\pm} \right) ,
\eea
The boundary terms arising on obtaining these equations are
\be \int d^{4}p \ph^2 \sqrt{-G} G^{zz}\pi^{a \ast}\left(z\right)\partial_z \pi^{a}\left(z\right)\mid_{z_{IR}}^{z_{UV}}=0 \label{83}
\ee
and the boundary conditions imposed on the $\pi^a$ fields are usual ones: $\partial_z \pi^{a}\left(z\right)\mid_{z_{IR}}=0$ and $\pi^{a}\left(z\right)\mid_{z_{UV}}=0$. Comparing the first equations with the second ones in (\ref{81}) and in (\ref{82}) one can establish following relations between the $\partial_z$ derivatives of the $\pi$ and $\chi$ fields
\begin{eqnarray}
\widetilde{\omega }_{0}^{2}G^{00}\partial_z\chi^0=4g^{2}\phi^{2}\partial_{z}\pi^{0}, \nonumber \\
\widetilde{\omega}_{\pm}\left(\widetilde{\omega}_{\pm}\mp V_0^3\right)G^{00} \partial_z\chi^{\pm} = 4g^{2}\phi^{2}\partial_{z}\pi^{\pm}.\label{84}
\end{eqnarray}
Taking derivative from the equations (\ref{81}) and (\ref{82}) and taking into account the relations (\ref{84}) the equation of motion for the $\pi^a$ field will be separated from the one for the $\chi^a$ field:
\begin{eqnarray}
\partial _{z}\left[\frac{1}{\phi^{2}\sqrt{-G}G^{00}}\partial_{z}\left( \phi^{2}\sqrt{-G}G^{zz}
\partial _{z}\pi^{0}\right)\right]-\left(\widetilde{\omega }_{0}^{2}+4g^{2}\phi^{2}G_{00}
\right)\partial_{z}\pi^{0}=0, \nonumber \\
\partial _{z}\left[\frac{1}{\phi^{2}\sqrt{-G}G^{00}}\partial_{z}\left( \phi^{2}\sqrt{-G}G^{zz}
\partial _{z}\pi^{\pm}\right)\right]-\left[\left(\widetilde{\omega}_{\pm}\mp V_{0}^{3}
\right)^{2}+4g^{2}\phi^{2}G_{00}\right]\partial_{z}\pi^{\pm}=0.
\label{85}
\end{eqnarray}
The physical situation with the $\pi$ fields is more complicated in comparison with vector and axial-vector fields considered above. As is seen from the action (\ref{57}) the $\pi$ fields interact with three other fields: the background condensate field $\phi$, the longitudinal axial-vector field $\chi$ and the background  isospin field $V_0^3$. In the action (\ref{57}) the interaction with the $\phi$ field gives a contribution to the mass term of the $\pi$ field and this contribution is the same for all $\pi^a$ components. The interaction with the isospin field $V_0^3$ gives different contributions to the mass terms of the different $\pi^a$ components in the equations of motion and, thus, splits the mass of these fields, which can be inferred from Eq. (\ref{85}). The interaction with the $\chi$ field lifts up the order in the equations of motion, which can be seen taking the $\chi$ field to be zero in the action (\ref{57}) or in Eqs. (81) and (82). In the case $\chi=0$ we have second order differential equations, instead of third order one in Eq. (\ref{85}). Another observation is that the $\phi(z)$ field modifies the AdS metric (\ref{11}). It is easy to see this in $\chi=0$ case of the equations for the $\pi$ fields in (81) and (82). Choosing following notation in the metric tensor:
\begin{equation}
G_{MM}=\phi^{-4/3}(z)\textbf{G}_{MM},\quad G^{MM}=\phi^{4/3}(z)\textbf{G}^{MM} \label{86}
\end{equation}
we get modified background geometry:
\begin{equation}
ds^{2}=\frac{R^{2}}{z^{2}\phi^{4/3}}\left( - dt^{2}+dz^{2}+d\overrightarrow{x}^{2}\right). \label{87}
\end{equation}
Then
\begin{equation}
\ph^2 \sqrt{-G} G^{zz}=\sqrt{-\textbf{G}}\textbf{G}^{zz},  \quad \ph^2 \sqrt{-G} G^{00}=  \sqrt{-\textbf{G}} \textbf{G}^{00}\label{88}
\end{equation}
and the equations for the $\pi$ fields in (81) and (82) get the form:
\begin{eqnarray}
-\o_0^2 \sqrt{-\textbf{G}} \textbf{G}^{00}  \pi^0 +\partial_z \left(\sqrt{-\textbf{G}} \textbf{G}^{zz} \partial_z  \pi^0 \right) =0, \nonumber\\
-\sqrt{-\textbf{G}} \textbf{G}^{00}\left( \o_{\pm} \mp V_0^3 \right)^2 \pi^{\pm} +  \pa_z \left( \sqrt{-\textbf{G}} \textbf{G}^{zz} \partial_z \pi^{\pm} \right)=0. \label{89}
\end{eqnarray}
As is seen from (\ref{89}) ignorance of the $\chi$ field in the action leads to the absence of the condensate term $4g^2\phi^2$ in equation of motion. This case is useful for a comparison with the mass splitting formula for pions obtained in \cite{34,47,48}. Equation (\ref{89}) is, in fact, the five dimensional Klein-Gordon equation for the free $\pi^a$ field with zero five dimensional mass term in the background geometry (\ref{87}):
\bea
\frac{1}{\sqrt{-\textbf{G}}}\partial_z \left(\sqrt{-\textbf{G}} \textbf{G}^{zz} \partial_z  \pi^0 \right)+\frac{1}{\sqrt{-\textbf{G}}}D_{\m}^{(a)}\left(\sqrt{-\textbf{G}}\textbf{G}^{\m{\m}^{\prime}}
D^{(a)}_{{\m}^{\prime}}\pi^a\right) \nonumber \\=\frac{1}{\sqrt{-\textbf{G}}}D_{M}^{(a)}\left(\sqrt{-\textbf{G}}\textbf{G}^{MM^{\prime}}D^
{(a)}_{M^{\prime}}
\pi^a\right)=0.\label{90}
\eea
 Consequently, the equation of motion for the $\pi^a$ field interacting with the $\phi$ field in the background geometry (\ref{11}) is equivalent to the one for the free $\pi^a$ field in the modified geometry (\ref{87}). It is obvious this equivalence does not depend on the isospin medium and is valid for the $V_0^3=0$ case as well. It should be noticed that the background geometry modification by scalar field $\phi$ was observed in \cite{56,57} also. Using (\ref{90}) we can derive a formula determining the effective mass of the $\pi$ fields. In order to make correspondence similar to previous sections let us write the Klein-Gordon equation in a four-dimensional curved space-time with the new metric $\textbf{G}$ for the massive scalar field $\pi^a$ interacting with the constant external gauge field $V_0^3$:
\begin{equation}
\frac{1}{\sqrt{-\textbf{G}}}D_{\m}^{(a)}\left(\sqrt{-\textbf{G}}\textbf{G}^{\m{\m}^{\prime}}
D^{(a)}_{{\m}^{\prime}}\pi^a\right)-\left(m^a\right)^2\pi^a=0,\label{91}
\end{equation}
where the $m^a$ is the mass of the $\pi^a$ field. If we are to consider Eq. (\ref{91}) as the UV limit of Eq. (\ref{90}) ones, then the eigenvalues of the operator $-\frac{1}{\sqrt{-\textbf{G}}}\partial_z\left(\sqrt{-\textbf{G}}\textbf{G}^{zz^{\prime}}
\partial_{z^{\prime}}\right)$ in (\ref{90}) will correspond to the squared mass $\left(m^a\right)^2$ of the 4D vector field $\pi^a$ in (\ref{91}). So, we may admit an equality of the eigenvalues of this operator to the effective mass of the $\pi$ field defined on this boundary:
\begin{equation}
-\frac{1}{\sqrt{-\textbf{G}}}\partial_z\left(\sqrt{-\textbf{G}}\textbf{G}^{zz^{\prime}}
\partial_{z^{\prime}}\pi^a\right)=\left(m^a\right)^2\pi^a_.\label{92}
\end{equation}
Equation (\ref{92}) is the one determining the effective mass of the $\pi$ field in the isospin medium by means of the eigenvalue of the derivative operator over the extra dimension.
For the Kaluza-Klein mode $\pi^a_s$  in the decomposition $$ \pi^a \left(x, z\right)=\sum_s b^{a s}(P^0)\pi_{s}^{a}(P^0, z)$$ a solution to Eq. (\ref{85}) is expressed in terms of the Bessel functions $J_1$ and $Y_1$:
\begin{equation}
\pi_{s}^{a}(z)=\frac{z}{\tilde{\textbf{m}}^{s}_{a}}\left[ c_{1}J_{1}\left(\tilde{\textbf{m}}^{s}_{a}z\right)
+c_{2}Y_{1}\left(\tilde{\textbf{m}}^{s}_{a}z\right) \right].
\label{93}
\end{equation}
Here the eigenvalues $\tilde{\textbf{m}}^{s}_{a}$ are the mass spectrum:
\begin{eqnarray}
\tilde{\textbf{m}}^{s}_{0}&=&\widetilde{\omega}^{s}_{0}\nonumber \\
\tilde{\textbf{m}}^{s}_{\pm}&=&\mid\widetilde{\omega}^{s}_{\pm}\mp V_0^3\mid=\mid\tilde{\textbf{m}}^{s}_{0}\mp V_0^3\mid \label{94}
\end{eqnarray}
and it has opposite splitting sign to the one in \cite{34,47,48}. For the solution (\ref{93}) the spectrum (\ref{43}) is available due to the Neumann boundary condition at the IR boundary and obviously, in this case it can be introduced usual additional boundary terms $S_{\pm}$ in order to avoid the $z_{IR}^{\pm}$ distinguishing from the $z_{IR}^0$ introduced in previous sections:
\be
S_{\pm}=\pm V_0^3\int d^4x \phi^2 \sqrt{-G} G^{zz} G^{mn} \left(\left(\pi^{\pm}_m\right)^{\ast}\pi^{\pm}_n\right)_{z=z_{IR}}= \pm V_0^3\int d^4x \sqrt{-\textbf{G}}\textbf{G}^{zz} \left(\left(\pi^{\pm \ m}\right)^{\ast}\pi^{\pm}_m\right)_{z=z_{IR}}.\label{95}
\ee

It is interesting to consider the $\chi \neq0$ case as well. In this case Eq. (\ref{85}) contains the condensate term $4g^2\phi^2$ and can be solved in the UV and IR limits of $z$. As is well known, in free case the equation of motion for the $\pi^a$ field was solved either in the chiral limit ($m_{q}=0$) \cite{58,59} or in the zero condensate limit ($\phi=0$). In our isospin medium case the equations (\ref{85}) will be solved in UV and IR limits of $z$, i.e. in $z\rightarrow 0$ and $z\rightarrow z_{IR}$ limits.
In order to find the UV limit solution we take $z\rightarrow 0$ limit from Eq. (\ref{85}) that leads to the usual Bessel equation for the $\partial_{z}\pi^{a}(z)$:
\begin{eqnarray}
&&\partial _{z}^{2}\left(\partial_{z}\pi^{0}\right)-\frac{1}{z}\partial _{z}\left(\partial_{z}
\pi^{0}\right)+\left[\widetilde{\omega }_{0}^{2}-g^{2}m_{q}^{2}+\frac{1}{z^{2}}\right]\left(\partial_{z}\pi^{0}\right)=0,\nonumber \\
&&\partial _{z}^{2}\left(\partial_{z}\pi^{\pm}\right)-\frac{1}{z}\partial _{z}\left(
\partial_{z}\pi^{\pm}\right)+\left[\left(\widetilde{\omega }_{\pm }\mp V^{3}_{0}\right)
^{2}-g^{2}m_{q}^{2}+\frac{1}{z^{2}}\right]\left(\partial_{z}\pi^{\pm}\right)=0.
\label{96}
\end{eqnarray}
Denoting $\partial_{z}\pi^{a}=zP^a(z)$ we shall get the Bessel equation for $P^a(z)$, which has a solution expressed in terms of the Bessel functions $J_0$ and $Y_0$ for the Kaluza-Klein mode $\pi_s^a$:
\begin{equation}
P_{s}^{a}(z)= c_{1}J_{0}\left(\tilde{\textbf{m}}^{s}_{a}z\right)+
c_{2}Y_{0}\left(\tilde{\textbf{m}}^{s}_{a}z\right). \label{97}
\end{equation}
Here the $\tilde{\textbf{m}}^{s}_{a}$ denote the square root of the constant in last terms in Eq. (\ref{91}):
\begin{eqnarray}
\tilde{\textbf{m}}^{s}_{0}&=&\sqrt{\left(\widetilde{\omega }_{0}^{s}\right)^{2}-
g^{2}m_{q}^{2}}, \nonumber \\
\tilde{\textbf{m}}^{s}_{\pm}&=&\sqrt{\left(\widetilde{\omega }^{s}_{\pm}\mp V_{0}^{3}\right)^{2}
-g^{2}m_{q}^{2}}. \label{98}
\end{eqnarray}
Using the differentiation formula for the Bessel functions
\be
\frac{d}{dx}\left[x^{\nu}Z_{\nu}\left(x\right)\right]=x^{\nu}Z_{\nu-1}\left(x\right)
\label{99}
\ee
we find the solution $\pi_{s}^a$ in terms of the Bessel functions $J_1$ and $Y_1$
\begin{equation}
\pi_{s}^{a}(z)=\frac{z}{\tilde{\textbf{m}}^{s}_{a}}\left[ c_{1}J_{1}\left(\tilde{\textbf{m}}^{s}_{a}z\right)
+c_{2}Y_{1}\left(\tilde{\textbf{m}}^{s}_{a}z\right) \right] .
\label{100}
\end{equation}
The boundary conditions which were imposed on the solution $\pi^a_s$ when deriving (\ref{85}) are usual ones: $\partial_z \pi_s^a \mid_{z_{IR}}=0$ and  $\pi^a_s \mid_{0}=0$. Similarly to the previous section, the application of UV boundary condition on (\ref{100}) gives $c_2=0$.

Now we solve the equations (\ref{85}) in the $z\rightarrow z_{IR}$ limit. We ignore $m_q$ in the condensate term due to $m_q\ll\sigma z_{IR}^2$. Making the substitution $\partial_{z}\pi^{a}=\frac{1}{z}P^{\prime a}\left(z\right)$ in these equations we obtain the following explicit form of them:
\begin{eqnarray}
&&\partial _{z}^{2}P^{\prime0}\left(z\right)+\frac{1}{z}\partial _{z}P^{\prime0}\left(z\right)+
\left[\widetilde{\omega }_{0}^{2}-g^{2}\sigma^2 z^4-\frac{4}{z^2}\right]P^{\prime0}\left(z\right)=0, \nonumber \\
&&\partial _{z}^{2}P^{\prime\pm}\left(z\right)+\frac{1}{z}\partial _{z}P^{\prime\pm}\left(z\right)+
\left[\left( \widetilde{\omega }_{\pm }\mp V^{3}_{0}\right) ^{2}-g^{2}\sigma^2 z^4
-\frac{4}{z^2}\right]P^{\prime\pm}\left(z\right)=0. \label{101}
\end{eqnarray}
In order to get the the IR asymptotic solution we shall take $z\rightarrow z_{IR}$ limit from the equations in (\ref{101}) and set the $z=z_{IR}$ in the condensate term. A numerical comparison of  terms $g^2\sigma^2 z^4$ and $1/z^2$ done when solving the Eq. (\ref{66}) is available for Eq. (\ref{101}) as well. As a result of this approximation we shall find the following Bessel function asymptotic solution for the equations in (\ref{101}):
\begin{equation}
P_{s}^{\prime a}(z)= c_{1}^{\prime}J_{2}\left(\tilde{\textbf{M}}^{s}_{a}z\right)+
c_{2}^{\prime}Y_{2}\left(\tilde{\textbf{M}}^{s}_{a}z\right), \label{102}
\end{equation}
where $\tilde{\textbf{M}}^{s}_{a}$ denote a square root of the value of brackets without the $4/z^2$ term in (\ref{101}) at $z_{IR}$:
\begin{eqnarray}
\tilde{\textbf{M}}^{s}_{0}&=&\sqrt{\left(\widetilde{\omega }_{0}^{s}\right)^{2}-g^{2}
\sigma^{2} z_{IR}^4}\approx\
\widetilde{\omega }_{0}^{s}-\frac{g^{2}\sigma^{2} z_{IR}^4}{2\widetilde{\omega }_{0}^{s}}, \nonumber \\
\tilde{\textbf{M}}^{s}_{\pm}&=&\sqrt{\left(\widetilde{\omega }^{s}_{\pm}\mp V_{0}^{3}\right)^{2}
-g^{2}\sigma^{2} z_{IR}^4}\approx \mid\widetilde{\omega }^{s}_{\pm}\mp V_{0}^{3}\mid
-\frac{g^{2}\sigma^{2} z_{IR}^4}
{2\mid\widetilde{\omega }^{s}_{\pm}\mp V_{0}^{3}\mid}. \label{103}
\end{eqnarray}
From (\ref{103}) we can write a mass splitting formula for this case:
\be
\tilde{\textbf{M}}^{s}_{\pm}\approx\mid \tilde{\textbf{M}}^{s}_{0}\mp V_{0}^{3}+ \frac{g^{2}\sigma^{2} z_{IR}^4}{2\widetilde{\omega}_{0}^{s}}\mid-\frac{g^{2}\sigma^{2} z_{IR}^4}{2\mid\widetilde{\omega }_{0}^s\mp V_0^3\mid}.
\label{104}
\ee
It is obvious  all three $\tilde{\textbf{M}}^{s}_{a}$ get same value at $V_0^3\rightarrow0$ limit.

Using (\ref{98}) and $Z_{-n}=(-1)^nZ_{n}$ property of the Bessel functions we can write the solutions $\pi^a_s$:
\begin{equation}
\pi_{s}^{a}(z)= -\frac{1}{z}\left[c_{1}^{\prime}J_{1}\left(\tilde{\textbf{M}}^{s}_{a}z\right)+
c_{2}^{\prime}Y_{1}\left(\tilde{\textbf{M}}^{s}_{a}z\right)\right]. \label{105}
\end{equation}
The UV boundary condition on this solution leads to $c_2=0$.
 In the UV limit, when $z\rightarrow0$, the $\sigma^2 z^4$ term drops and the solution to  (\ref{85}) will be same as the one for Eq. (\ref{66}). The general solution to Eq. (\ref{85}) will be a linear combination of these asymptotic solutions. It can be shown the constants $c_2$ and $c_2^{\prime}$ are zero in this case as well and the general solution has got the form:
\begin{equation}
\pi_{s}^{a}(z)= c_{1}zJ_{1}\left(\tilde{\textbf{m}}^{s}_{a}z\right)+
\frac{c_{1}^{\prime}}{z}J_{1}\left(\tilde{\textbf{M}}^{s}_{a}z\right). \label{106}
\end{equation}
We shall apply the Neumann boundary condition at the IR boundary only on the IR asymptotic part solution  (\ref{106}),  i.e. only on the second term in (\ref{106}). This gives us a discrete mass spectrum in terms of the zeros $\alpha^s_2$ of the Bessel function $J_{2}$:
\begin{eqnarray}
\tilde{\textbf{M}}^{s}_{0}&=&\frac{\alpha^s_2}{\tilde{z}_{IR}^{0}},\nonumber \\
\tilde{\textbf{M}}^{s}_{\pm }&=&\frac{\alpha^s_2}{\tilde{z}_{IR}^{\pm}};\quad s=1,2,3...
\label{107}
\end{eqnarray}
In order to verify our consideration of $\tilde{\textbf{M}}^{s}_{a}$ as an effective mass we should reduce Eq. (\ref{85}) to the form in Eq. (\ref{90}) in the UV limit  $z\rightarrow 0$. Let us write Eq. (\ref{85}) in terms of induced metric (\ref{86}):
\begin{equation}
\partial_z\left[\frac{1}{\sqrt{-\textbf{G}}\textbf{G}^{00}}\partial_z\left(
\sqrt{-\textbf{G}}\textbf{G}^{zz}\partial_z \pi^a\right)+\left(D_{0}^aD^a_0+4g^2\phi^{5/2}\textbf{G}_{00}\right)\pi^a\right]=4g^2\pi^a\partial_z\left(\phi^{5/2}
\textbf{G}_{00}\right).\label{108}
\end{equation}
The right hand side of this equation in the UV limit gives zero due to UV boundary condition $\pi^a\mid_0=0$. Then the free constant obtained in the left hand side can be chosen zero and (\ref{108}) will be written as:
\begin{equation}
\frac{1}{\sqrt{-\textbf{G}}}\partial_z\left(\sqrt{-\textbf{G}}\textbf{G}^{zz}
\partial_z \pi^a\right)+\textbf{G}^{00}\left(D_{0}^aD^a_0+4g^2\phi^{5/2}\textbf{G}_{00}\right)\pi^a=0.
\label{109}
\end{equation}
This equation has a form of the the equations in (\ref{89}) and can be considered as a Klein-Gordon equation in the background geometry with the metric $\textbf{G}_{MN}$.  Then second term will be considered as an effective mass including the condensate contribution. This proves that the derivative operator over the extra dimension $z$ determines the effective mass through eigenvalues $\tilde{\textbf{M}}^{s}_{a}$ in this non-zero condensate term case as well.

In this case the additional boundary terms compensating for the $z_{IR}$ distinguished will be dependent on the condensate $g^{2}\sigma^{2} z_{IR}^4$ and the state mass $\widetilde{\omega }^{s}_{a}$:
\bea
S_{\pm}=\sum_s\left(\pm V_0^3+\frac{g^{2}\sigma^{2} z_{IR}^4}{2\mid\widetilde{\omega }_{0}^s\mp V_0^3\mid}\right)\int d^4x \phi^2 \sqrt{-G} G^{zz}\left(b^{\pm s}\right)^2\left(\pi^{\pm}_s\right)^2_{z=z_{IR}}\nonumber \\ =\sum_s \left(\pm V_0^3+\frac{g^{2}\sigma^{2} z_{IR}^4}{2\mid\widetilde{\omega }_{0}^s\mp V_0^3\mid}\right)\int d^4x \sqrt{-\textbf{G}}\textbf{G}^{zz}\left(b^{\pm s}\right)^2\left(\pi^{\pm}_s\right)^2_{z=z_{IR}} \nonumber \\
S_{0}=\sum_s \frac{g^{2}\sigma^{2} z_{IR}^4}{2\widetilde{\omega }_{0}^s}\int d^4x \sqrt{-\textbf{G}}\textbf{G}^{zz} \left(b^{0s}\right)^2\left(\pi^{0}_s\right)^2_{z=z_{IR}}.\label{110}
\eea
Since the condensate term $4g^2\phi^2$ shifts masses of all pions, the boundary terms above include it also.
\section{Discussion}
Here we obtain a formula for the meson's effective mass in the isospin medium using holography. We observe that interaction with the isospin background leads to the necessity of distinguishing  of the infrared boundaries for different mesons from the same isotriplet. To bring all infrared boundaries to the same value requires one to introduce additional boundary terms, which leads to a modification of boundary  condition in the presence of an isospin background field. Indeed, the  modification of the boundary condition at infrared boundary  is known from the works \cite{37,52}. In \cite{52}, the axial-vector field interacting with condensate field gets a contribution in mass and this results in a modification in the boundary condition at $z_{IR}$. The shift in the boundary condition which was made in this work is equivalent to adding new boundary term proportional to the condensate field. Here we have an isospin field interaction with the $\rho$ mesons, which contributes to the mass differently for the $\rho^+$ and $\rho^-$ mesons. This results in the introduction of different boundary terms for these mesons. In the case of $a_1$ and $\pi$ mesons we have two fields (condensate field and isospin field), the interaction with which contributes to the mass of these mesons. In the result we have double shift in the mass and this requires one to introduce boundary terms proportional to both these fields.

{\bf Acknowledgements}

The author thanks Joshua Erlich for useful discussion of adding new boundary terms and for commenting on the manuscript. He acknowledges Chanyong Park and Bum-Hoon Lee for useful discussions at the beginning stage of this work. Author thanks TUBITAK organization of Turkey for the grant 2221 - Fellowships for Visiting Scientists and Scientists on Sabbatical Leave and Physics Department of Gazi University, where the main part of present work was done.
 for the hospitality during visiting.


\begin{thebibliography}{99}
\bibitem{1}
  D.~T.~Son and M.~A.~Stephanov,
  Phys.\ Rev.\ Lett.\  {\bf 86}, 592 (2001)
  [arXiv:0005225 [hep-ph]].

\bibitem{2}
  J.~M.~Maldacena,
  Adv.\ Theor.\ Math.\ Phys.\  {\bf 2}, 231 (1998)

\bibitem{3}
  J.~M.~Maldacena,
  [Int.\ J.\ Theor.\ Phys.\  {\bf 38}, 1113 (1999)]
  [arXiv:hep-th/9711200].

\bibitem{4}
  S.~S.~Gubser, I.~R.~Klebanov and A.~M.~Polyakov,
  Phys.\ Lett.\ B {\bf 428}, 105 (1998)
  [hep-th/9802109].

\bibitem{5}
  E.~Witten,
  Adv.\ Theor.\ Math.\ Phys.\  {\bf 2}, 253 (1998)
  [hep-th/9802150].

\bibitem{6}
  O.~Aharony, S.~S.~Gubser, J.~M.~Maldacena, H.~Ooguri and Y.~Oz,
  Phys.\ Rept.\  {\bf 323}, 183 (2000)
  [hep-th/9905111].

\bibitem{7}  H. Boschi-Filho and N.R.F. Braga, JHEP {\bf 0305} (2003) 009
 [arXiv:0212207[hep-th]]

\bibitem{8}  H. Boschi-Filho and N.R.F. Braga, Eur.\ Phys.\ J. C {\bf 32} (2004) 529, [arXiv: 0209080[hep-th]]

\bibitem{9}
  J.~Erlich, E.~Katz, D.~T.~Son and M.~A.~Stephanov,
  Phys.\ Rev.\ Lett.\  {\bf 95}, 261602 (2005)
  [hep-ph/0501128].

\bibitem{10}
  A.~Karch, E.~Katz, D.~T.~Son and M.~A.~Stephanov,
  Phys.\ Rev.\ D {\bf 74}, 015005 (2006)
  [hep-ph/0602229].

\bibitem{11}
  L.~Da Rold and A.~Pomarol,
  Nucl.\ Phys.\ B {\bf 721}, 79 (2005)
  [hep-ph/0501218].

\bibitem{12} L.~Da Rold and A.~Pomarol,
  JHEP {\bf 0601}, 157 (2006)
  [hep-ph/0510268].
 \bibitem{13}
S. J. Brodsky, G. F. de Teramond, H. G. Dosch and J. Erlich, "Light-Front Holographic QCD and Emerging Confinement", [arXiv:1407.8131 [hep-ph]].

\bibitem{14} J. Erlich, Contemp. Phys. {\bf 56}, 159 (2015). arXiv:1407.5002 [hep1068
ph]

\bibitem{15}
  D.~K.~Hong, T.~Inami and H.~-U.~Yee,
  Phys.\ Lett.\ B {\bf 646}, 165 (2007)
  [hep-ph/0609270].

\bibitem{16} Z. Abidin and C.Carlson, Phys. Rev. D {\bf 79}, 115003 (2009),[arXiv:0903.4818[hep-ph]]
\bibitem{17}
E. Witten,  Adv.Theor.Math.Phys. {\bf 2}, 505 (1998), [arXiv:9803131 [hep-th]]

\bibitem{18}
  K.~Ghoroku and M.~Yahiro,
  Phys.\ Rev.\ D {\bf 73}, 125010 (2006)
  [hep-ph/0512289].
\bibitem{19}
 P.~Colangelo, F.~Giannuzzi, S.~Nicotri and V.~Tangorra,
  Eur.\ Phys.\ J.\ C {\bf 72}, 2096 (2012)
  [arXiv:1112.4402 [hep-ph]];

\bibitem{20}
P.~Colangelo, F.~Giannuzzi and S.~Nicotri,
  JHEP {\bf 1205}, 076 (2012) arXiv:1201.1564 [hep-ph]].
\bibitem{21}
M. Fujita, K. Fukushima, T. Misumi and M. Murata,
 Phys.Rev. D {\bf 80} (2009) 035001
[arXiv:0903.2316 [hep-ph]].

\bibitem{22}
A.S. Miranda, C.A. Ballon Bayona, H. Boschi-Filho and N. R.F. Braga
JHEP {\bf 0911} (2009) 119 [arXiv:0909.1790 [hep-th]].

\bibitem{23}
L. A.H. Mamani, A. S. Miranda, H. Boschi-Filho and N. R. F. Braga,
JHEP {\bf 1403} (2014) 058, [arXiv:1312.3815 [hep-th]].

\bibitem{24}	
A. S. Miranda, C.A. Ballon Bayona, H. Boschi-Filho and Nelson R.F. Braga
Nucl. Phys. Proc. Suppl. 199 (2010) 107-112, [arXiv:0910.4319 [hep-th]].

\bibitem{25}
Z. Li and B.-Q. Ma, Phys. Rev. D {\bf 89} (2014) 015014
 [arXiv:1312.3451 [hep-ph]].

\bibitem{26}
  C.~Park,
  Phys.\ Rev.\ D {\bf 81}, 045009 (2010)
  [arXiv:0907.0064 [hep-ph]].


\bibitem{27}
 J. Casalderrey-Solana, H. Liu, D. Mateos, K. Rajagopal, U.A. Wiedemann, Gauge/string duality, hot QCD and heavy ion collisions (Cambridge University Press, Cambridge, 2014)  [arXiv:1101.0618 [hep-th]].

\bibitem {28}
E. Papantonopoulos, From Gravity to Thermal Gauge Theories:
The AdS/CFT Correspondence. Lecture Notes in Physics, vol.{\bf 828}
(Springer, Berlin, 2011).


 \bibitem{29}
 I.Ya Aref'eva, Phys.Usp. {\bf 57} (2014) 527

 \bibitem{30}
  A.~Parnachev,
  JHEP {\bf 0802}, 062 (2008)
  [arXiv:0708.3170 [hep-th]].

\bibitem{31}
  O.~Aharony, K.~Peeters, J.~Sonnenschein and M.~Zamaklar,
  JHEP {\bf 0802}, 071 (2008)
  [arXiv:0709.3948 [hep-th]].
\bibitem{32}
N. Evans, K.-Y. Kim, M. Magou, Y. Seo and S.-J. Sin, JHEP {\bf 1209} (2012) 045
[arXiv:1204.5640 [hep-th]].
\bibitem{33} 
Y. Seo, J.P. Shock, S.-J. Sin, D. Zoakos, JHEP {\bf 1003} (2010) 115
[arXiv:0912.4013 [hep-th]].
\bibitem{34}
  D.~Albrecht and J.~Erlich,
  Phys.\ Rev.\  D {\bf 82}, 095002 (2010).
  arXiv:1007.3431 [hep-ph].
\bibitem{35} K.~-I.~Kim, Y.~Kim and S.~H.~Lee, J. Korean Phys. Soc. 55, 1381 (2009).
  arXiv:0709.1772 [hep-ph].

\bibitem{36} S.S. Afonin, A.A. Andrianov, D. Espriu, Phys.Lett. B {\bf 745} (2015) 52.
arXiv:1503.08606 [hep-ph]
\bibitem{37}
D. Albrecht, J. Erlich and R.J. Wilcox, Phys.Rev. D {\bf 85}, 114012 (2012).
arXiv:1112.5643 [hep-ph]
\bibitem{38}
H. Nishihara and M. Harada, Phys.Rev. D{\bf 89}, 076001 (2014)
arXiv:1401.2928 [hep-ph]
\bibitem {39}
 H. Nishihara, M. Harada, Phys. Rev. D {\bf 90}, 115027 (2014).
arXiv:1407.7344 [hep-ph]
\bibitem {40} C.~P.~Herzog,
  Phys.\ Rev.\ Lett.\  {\bf 98}, 091601 (2007)
  arXiv:0608151 [hep-th].

\bibitem{41}
C.A. Ballon Bayona, H. Boschi-Filho, N.R.F. Braga and L.A. Pando Zayas, Phys.Rev. D{\bf77} 046002  (2008).  arXiv:0705.1529 [hep-th]
\bibitem{42}
  B.~-H.~Lee, C.~Park and S.~-J.~Sin,
  JHEP {\bf 0907}, 087 (2009).  arXiv:0905.2800 [hep-th].

\bibitem{43}
  K.~Jo, B.~-H.~Lee, C.~Park and S.~-J.~Sin,
  JHEP {\bf 1006}, 022 (2010). arXiv:0909.3914 [hep-ph].

\bibitem{44}  P.~Colangelo, F.~Giannuzzi and S.~Nicotri,
  Phys.\ Rev.\ D {\bf 83}, 035015 (2011).
  arXiv:1008.3116 [hep-ph].

\bibitem{45} C.~Park, D.~-Y.~Gwak, B.~-H.~Lee, Y.~Ko and S.~Shin,
  Phys.\ Rev.\ D {\bf 84}, 046007 (2011)
  [arXiv:1104.4182 [hep-th]];

\bibitem{46}
  Y.~Kim, B.~-H.~Lee, S.~Nam, C.~Park and S.~-J.~Sin,
  Phys.\ Rev.\ D {\bf 76}, 086003 (2007)
  [arXiv:0706.2525 [hep-ph]].

\bibitem{47}B.-H. Lee, Sh. Mamedov, S. Nam and C. Park, JHEP {\bf 1308} (2013) 045
\bibitem{48}B.-H. Lee, Sh. Mamedov and C. Park, Int. J. Mod. Phys. A {\bf29}, 1450170 (2014) [arXiv: hep-th/1402.6061].
\bibitem{49}
B.-H. Lee, C. Park and S. Nam,  JHEP 1505 (2015) 011
[arXiv:1412.3097 [hep-ph]]
\bibitem{50}
B.-H. Lee and C. Park,  Phys.Lett. B {\bf 746} (2015) 202
[arXiv:1503.03615 [hep-th]]
\bibitem{51}	
Y. Kim, Y. Seo, I.J. Shin, S.-J. Sin, Holographic Meson Mass in Asymmetric Dense Matter,(2011)
 arXiv:1108.2751 [hep-ph]
\bibitem{52}
 N. Maru and M. Tachibana,
Eur. Phys. J. C {\bf 63}, 123 (2009). arXiv:0904.3816 [hep-ph]
\bibitem{53}
  C.~Park,
  Phys.\ Lett.\ B {\bf 708}, 324 (2012)
  arXiv:1112.0386 [hep-th].
\bibitem{54} H.R. Grigoryan and A.V. Radyushkin, Phys. Rev. D {\bf 76}, 115007 (2007).  arXiv:0709.0500 [hep-ph].

\bibitem{55} A. Mammarella and M. Mannarelli, Phys. Rev. D{\bf 92} 085025 (2015), arxiv: 1507.02934 [hep-ph]
\bibitem{56}
  B.~-H.~Lee, C.~Park and S.~Shin,
  JHEP {\bf 1012}, 071 (2010).   arXiv:1010.1109 [hep-th].
\bibitem{57}
  C.~Park, B.~-H.~Lee and S.~Shin,
  Phys.\ Rev.\ D {\bf 85}, 106005 (2012)
  [arXiv:1112.2177 [hep-th]].

\bibitem{58} H.R. Grigoryan and A.V. Radyushkin, Phys. Lett. B {\bf 650}, 421 (2007) [arXiv: hep-ph/0703069]


\bibitem{59} H.C. Kim, Y.Kim and U. Yakhshiev, JHEP {\bf 0911}, 034 (2009) arXiv:
hep-ph/ 0908.3406


\end{thebibliography}
\end{document}